\documentclass[]{pasj01}
\usepackage{lscape}
\newcommand{\pasa}{PASA}

\begin{document} 
\Received{}
\Accepted{}

\title{ALMA TWENTY-SIX ARCMIN$^2$ SURVEY OF GOODS-S AT ONE-MILLIMETER (ASAGAO): Source Catalog and Number Counts}

\author{
Bunyo \textsc{Hatsukade},\altaffilmark{1}
Kotaro~\textsc{Kohno},\altaffilmark{1,2}
Yuki~\textsc{Yamaguchi},\altaffilmark{1}
Hideki~\textsc{Umehata},\altaffilmark{1,3}
Yiping~\textsc{Ao},\altaffilmark{4}
Itziar~\textsc{Aretxaga},\altaffilmark{5}
Karina~I.~\textsc{Caputi},\altaffilmark{6}
James~S.~\textsc{Dunlop},\altaffilmark{7}
Eiichi~\textsc{Egami},\altaffilmark{8}
Daniel~\textsc{Espada},\altaffilmark{9,10}
Seiji~\textsc{Fujimoto},\altaffilmark{11}
Natsuki~\textsc{Hayatsu},\altaffilmark{12,13}
David~H.~\textsc{Hughes},\altaffilmark{5}
Soh~\textsc{Ikarashi},\altaffilmark{6}
Daisuke~\textsc{Iono},\altaffilmark{9,10}
Rob~J.~\textsc{Ivison},\altaffilmark{13,7}
Ryohei~\textsc{Kawabe},\altaffilmark{9,10}
Tadayuki~\textsc{Kodama},\altaffilmark{14}
Minju~\textsc{Lee},\altaffilmark{15}
Yuichi~\textsc{Matsuda},\altaffilmark{9,10}
Kouichiro~\textsc{Nakanishi},\altaffilmark{9,10}
Kouji~\textsc{Ohta},\altaffilmark{16}
Masami~\textsc{Ouchi},\altaffilmark{11,17}
Wiphu~\textsc{Rujopakarn},\altaffilmark{17,18,19}
Tomoko~\textsc{Suzuki},\altaffilmark{9}
Yoichi~\textsc{Tamura},\altaffilmark{15}
Yoshihiro~\textsc{Ueda},\altaffilmark{16}
Tao~\textsc{Wang},\altaffilmark{1,9}
Wei-Hao~\textsc{Wang},\altaffilmark{20}
Grant~W.~\textsc{Wilson},\altaffilmark{21}
Yuki~\textsc{Yoshimura},\altaffilmark{1}
\and Min~S.~\textsc{Yun}\altaffilmark{21}
}

\altaffiltext{1}{Institute of Astronomy, Graduate School of Science, The University of Tokyo, 2-21-1 Osawa, Mitaka, Tokyo 181-0015, Japan}
\email{hatsukade@ioa.s.u-tokyo.ac.jp}
\altaffiltext{2}{Research Center for the Early Universe, The University of Tokyo, 7-3-1 Hongo, Bunkyo, Tokyo 113-0033, Japan}
\altaffiltext{3}{RIKEN Cluster for Pioneering Research, 2-1 Hirosawa, Wako-shi, Saitama 351-0198, Japan}
\altaffiltext{4}{Purple Mountain Observatory \& Key Laboratory for Radio Astronomy, Chinese Academy of Sciences, 8 Yuanhua Road, Nanjing 210034, China}
\altaffiltext{5}{Instituto Nacional de Astrof\'{\i}sica, \'{O}ptica y Electr\'{o}nica (INAOE), Luis Enrique Erro 1, Sta. Ma. Tonantzintla, Puebla, Mexico}
\altaffiltext{6}{Kapteyn Astronomical Institute, University of Groningen, P.O. Box 800, 9700AV Groningen, The Netherlands}
\altaffiltext{7}{Institute for Astronomy, University of Edinburgh, Royal Observatory, Edinburgh EH9 3HJ UK}
\altaffiltext{8}{Steward Observatory, University of Arizona, 933 N. Cherry Ave, Tucson, AZ 85721, USA}
\altaffiltext{9}{National Astronomical Observatory of Japan, 2-21-1 Osawa, Mitaka, Tokyo 181-8588, Japan}
\altaffiltext{10}{SOKENDAI (The Graduate University for Advanced Studies), 2-21-1 Osawa, Mitaka, Tokyo 181-8588, Japan}
\altaffiltext{11}{Institute for Cosmic Ray Research, The University of Tokyo, Kashiwa, Chiba 277-8582, Japan}
\altaffiltext{12}{Department of Physics, Graduate School of Science, The University of Tokyo, 7-3-1 Hongo, Bunkyo, Tokyo, 113-0033, Japan}
\altaffiltext{13}{European Southern Observatory, Karl-Schwarzschild-Str. 2, D-85748 Garching, Germany}
\altaffiltext{14}{Astronomical Institute, Tohoku University, Aramaki, Aoba-ku, Sendai, Miyagi 980-8578, Japan}
\altaffiltext{15}{Department of Physics, Nagoya University, Furo-cho, Chikusa-ku, Nagoya 464-8601, Japan}
\altaffiltext{16}{Department of Astronomy, Kyoto University, Kyoto 606-8502, Japan}
\altaffiltext{17}{Kavli Institute for the Physics and Mathematics of the Universe, Todai Institutes for Advanced Study, the University of Tokyo, Kashiwa, Japan 277-8583 (Kavli IPMU, WPI)}
\altaffiltext{18}{Department of Physics, Faculty of Science, Chulalongkorn University, 254 Phayathai Road, Pathumwan, Bangkok 10330, Thailand}
\altaffiltext{19}{National Astronomical Research Institute of Thailand (Public Organization), Don Kaeo, Mae Rim, Chiang Mai 50180, Thailand}
\altaffiltext{20}{Institute of Astronomy and Astrophysics, Academia Sinica, Taipei, Taiwan}
\altaffiltext{21}{Department of astronomy, University of Massachusetts, Amherst, MA 01003, USA}


\maketitle

\begin{abstract}
We present the survey design, data reduction, construction of images, and source catalog of the Atacama Large Millimeter/submillimeter Array (ALMA) twenty-six arcmin$^2$ survey of GOODS-S at one-millimeter (ASAGAO). 
ASAGAO is a deep ($1\sigma \sim 61$~$\mu$Jy~beam$^{-1}$ for a 250 k$\lambda$-tapered map with a synthesized beam size of $0\farcs51 \times 0\farcs45$) and wide area (26~arcmin$^2$) survey on a contiguous field at 1.2~mm. 
By combining with ALMA archival data in the GOODS-South field, we obtained a deeper map in the same region ($1\sigma \sim 30$~$\mu$Jy~beam$^{-1}$ for a deep region with a 250 k$\lambda$-taper, and a synthesized beam size of $0\farcs59 \times 0\farcs53$), providing the largest sample of sources (25 sources at $\ge$5.0$\sigma$, 45 sources at $\ge$4.5$\sigma$) among ALMA blank-field surveys to date. 
The number counts shows that $52^{+11}_{-8}$\% of the extragalactic background light at 1.2~mm is resolved into discrete sources at $S_{\rm 1.2mm} > 135$~$\mu$Jy. 
We create infrared (IR) luminosity functions (LFs) in the redshift range of $z =$ 1--3 from the ASAGAO sources with $K_S$-band counterparts, and constrain the faintest luminosity of the LF at $2.0 < z < 3.0$. 
The LFs are consistent with previous results based on other ALMA and SCUBA-2 observations, which suggest a positive luminosity evolution and negative density evolution with increasing redshift. 
We find that obscured star-formation of sources with IR luminosities of $\log{(L_{\rm IR}/L_{\odot})} \gtrsim 11.8$ account for $\approx$60\%--90\% of the $z \sim 2$ cosmic star-formation rate density. 
\end{abstract}

\section{Introduction} \label{sec:introduction}

Revealing cosmic star formation history is one of the biggest challenges in astronomy. 
Because a significant fraction of star formation is obscured by dust at high redshift (e.g., \cite{mada14}, for a review), infrared (IR)--submillimeter/millimeter (submm/mm) observations are required to understand the true star-forming activity. 
The intensity of the extragalactic background light (EBL) in the IR--submm/mm is known to be comparable to that of the EBL in the optical, also showing the importance of IR--submm/mm observations for revealing the dust-obscured activity in the Universe. 
Deep surveys at submm/mm (850~$\mu$m and 1~mm wavelengths) with ground-based telescopes uncovered a population of bright ($S_{\rm 1mm} \gtrsim 1$ mJy) submm/mm galaxies (SMGs; \cite{blai02, case14}, for reviews). 
SMGs are highly obscured by dust, and the resulting thermal dust emission dominates the bolometric luminosity. 
The energy source of submm/mm emission is primarily from intense star formation activity, with IR luminosities of $L_{\rm IR} \gtrsim$ a few $\times 10^{12}$~$L_{\odot}$ and star formation rates of SFRs $\gtrsim$ a few $\times 100 M_{\odot}$~yr$^{-1}$. 
The redshift distribution of SMGs is characterized by a median redshift of $z \sim 2$--3 (e.g., \cite{chap05, yun12, simp14, chen16, mich17, bris17}). 
The stellar masses and SFRs of SMGs show that they are located above or at the massive end of the main sequence of star-forming galaxies (e.g., \cite{dadd07, mich12, mich14, dacu15}). 
It is thought that SMGs are progenitors of massive elliptical galaxies in the present-day Universe observed during their formation phase (e.g., \cite{lill96, smai04}). 
The contribution of SMGs to the EBL is estimated by integrating the number counts. 
Blank field surveys with single-dish telescopes resolved $\sim$20\%--40\% of the EBL at 850~$\mu$m (e.g., \cite{barg99, eale00, bory03, copp06}) and $\sim$10\%--20\% at 1 mm (e.g., \cite{grev04, pere08, scot08, scot10, hats11}). 
It is expected that deeper submm/mm observations trace less dust-obscured star-forming galaxies, which may overlap galaxies detected in rest-frame ultraviolet (UV) and optical wavelengths. 
\citet{whit17} found a dependence of the fraction of obscured star formation (SFR$_{\rm IR}$) on stellar mass out to $z = 2.5$: 
50\% of star formation is obscured for galaxies with $\log(M/M_{\odot}) = 9.4$, and $>$90\% for galaxies with $\log(M/M_{\odot}) > 10.5$. 
Deep surveys probing fainter submm objects ($S_{\rm 1mm} < 1$~mJy), which are expected to be more normal star-forming galaxies rather than ``classical'' SMGs, are essential to understand the cosmic star-formation history and the origin of EBL, however, such observations have been hampered by the confusion limit of observations with single-dish telescopes since they have large beam sizes ($\sim$$15''$--$30''$).

Interferometric observations enable us to reveal faint submm sources by substantially reducing the confusion limit. 
The Atacama Large Millimeter/submillimeter Array (ALMA) is now detecting submm sources more than an order of magnitude fainter than ``classical'' SMGs. 
Because of its high sensitivity and high angular resolution, ALMA can collect serendipitous sources from a variety of data sets to probe the fainter end of the number counts (\cite{hats13, ono14, carn15, fuji16, oteo16}).
These studies show that more than 50\% of the EBL at 1 mm is resolved into discrete sources at a flux limit of $\sim$0.1~mJy.

These studies are based on serendipitous sources detected in fields where faint submm sources are not the main targets, which could introduce biases due to the clustering of sources around the targets or sidelobes caused by bright targets. 
It is necessary to conduct ``unbiased'' surveys in a contiguous field rather than collecting discrete fields in order to obtain a census on the population of faint submm sources. 
Surveys in a contiguous field are also beneficial for clustering analysis. 
During ALMA Cycle 1, the central 2 arcmin$^2$ area of the Subaru/XMM-Newton Deep Survey Field (SXDF) was observed as an ALMA deep blank field survey \citep{kohn16, tada15, hats16, wang16, yama16}. 
From Cycle 1 to present, the GOODS-S/Hubble Ultra Deep Field (HUDF) has been observed with ALMA in different surveys \citep{walt16, arav16, dunl17, fran18}. 
There are also deep surveys in overdense regions such as the ALMA deep field in the $z = 3.09$ protocluster SSA~22 field (ADF22; \cite{umeh15, umeh17, umeh18}) and the ALMA Frontier Fields Survey of gravitational lensing clusters \citep{gonz17}.

The GOODS-S/HUDF field has the deepest multi-wavelengths data from X-ray to radio with ground-based telescopes and satellites such as 
{\sl Chandra} \citep{xue11, luo17}, 
{\sl XMM-Newton} \citep{coma11}, 
{\sl HST}/ACS/WFC3 (HUDF, CANDELS, XDF; \cite{beck06, grog11, koek11, elli13, illi13}), 
VLT/HAWK-I (HUGS; \cite{font14}), 
Magellan/FourStar (ZFOURGE; \cite{stra16}), 
{\sl Spitzer} (S-CANDELS; \cite{ashb15}), 
{\sl Herschel}/PACS (PEP; \cite{lutz11}) and SPIRE (HerMES; \cite{oliv12}), 
APEX/LABOCA (LESS; \cite{weis09}), 
ASTE/AzTEC \citep{scot10, yun12}, 
SCUBA-2/JCMT \citep{cowi17}, 
and VLA \citep{mill13, rujo16}. 
Spectroscopic observations have also been conducted extensively (e.g., \cite{lefe04, bram12, skel14}). 
The VLT/MUSE spectroscopic survey of HUDF (the $3' \times 3'$ deep region region and $1' \times 1'$ ultra-deep region) provides 3-D data cubes of this field \citep{baco15, baco17}. 
JWST will conduct deep multi-band imaging and spectroscopy, offering the ability to diagnose optically-faint galaxies which are difficult to study with existing optical/near-IR telescopes.

\begin{figure}
\begin{center}
\includegraphics[width=\linewidth]{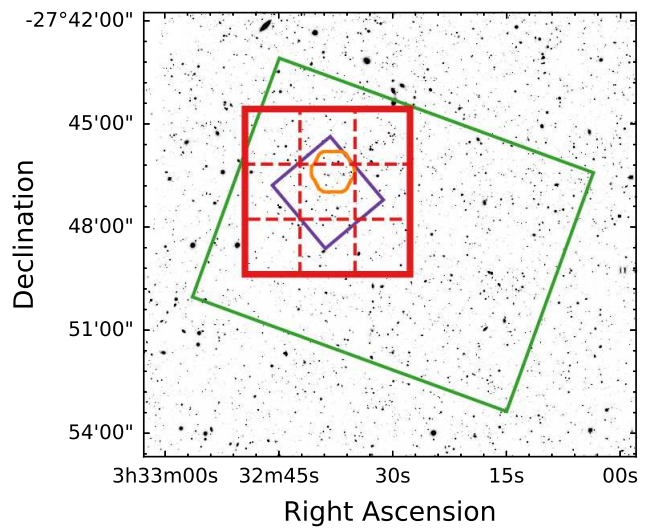}
\end{center}
\caption{
ASAGAO region consisting of nine sub-regions (red) overlaid on the {\sl HST}/WFC3 F160W image. 
The orange, purple, and green regions represent the ALMA survey areas of ASPECS \citep{walt16, arav16} at 1.2~mm, HUDF \citep{dunl17} at 1.3~mm, and GOODS-ALMA \citep{fran18} at 1.1~mm, respectively. 
}
\label{fig:region}
\end{figure}

The ALMA surveys of the GOODS-S field have been conducted with different survey strategies: 
a deep but narrow survey (4.5~arcmin$^2$, $1\sigma = 34$~$\mu$Jy~beam$^{-1}$) at 1.3 mm (HUDF; \cite{dunl17}), 
a shallower and wider survey (69~arcmin$^2$, $1\sigma \sim 180$~$\mu$Jy~beam$^{-1}$) at 1.1 mm (GOODS-ALMA; \cite{fran18}), 
and spectral scans in an area of 1~arcmin$^2$ (ALMA Spectroscopic Survey; ASPECS) at 3 mm and 1.2 mm \citep{walt16, arav16} (figure~\ref{fig:region}). 
The spectral scans cover the full window of the bands, offering the deepest continuum maps ($1\sigma_{\rm 3mm} = 3.8$~$\mu$Jy~beam$^{-1}$ and $1\sigma_{\rm 1.2mm} = 12.7$~$\mu$Jy~beam$^{-1}$).

The faint submm sources detected in these studies are found to be on the main sequence, but located at higher stellar mass and SFR ranges (e.g., \cite{hats15, yama16, arav16, dunl17}) due to the survey detection limit. 
In addition, the numbers of sources studied in these surveys are still very limited, 
and the demand for deeper and wider surveys remains high. 
In this paper, we present the results of ALMA twenty-six arcmin$^2$ survey of GOODS-S at one-millimeter (ASAGAO). 
ASAGAO is a deep ($1\sigma \sim 61$~$\mu$Jy~beam$^{-1}$ for a 250 k$\lambda$-tapered map) and wide-area (26~arcmin$^2$) survey on a contiguous field at 1.2~mm. 
The observing area matches the deepest VLA C-band 5 cm (6 GHz) observations (\cite{rujo16}; Rujopakarn et al. in prep.) and the ultra-deep VLT/HAWK-I $K_S$-band images. 
The primary goal of this survey is to obtain a census of galaxies with $L_{\rm IR} \gtrsim 3 \times 10^{11}$~$L_{\odot}$ or SFR $\gtrsim 50$~$M_{\odot}$~yr$^{-1}$ for the understanding of the dust-obscured star-formation history of the Universe. 
The initial results based on the ASAGAO data have been reported by \citet{ueda18} for the X-ray active galactic nucleus (AGN) properties, and by \citet{fuji18} for morphological studies. 
The results of the multi-wavelength analysis are discussed in \citet{yama18}, and the clustering analysis is conducted by Yoshimura et al. (in prep.).

The arrangement of this paper is as follows. 
Section~\ref{sec:data} outlines the ALMA observations, data reduction, and archival data used in this study, and shows the obtained images. 
Section~\ref{sec:source} describes the detected sources, and we list the source catalog. 
In Section~\ref{sec:counts}, we describe the method of creating number counts, and compare with previous studies. 
We present the method of constructing luminosity functions and compare with previous studies in Section~\ref{sec:lf}.  
The conclusions are presented in Section~\ref{sec:conclusions}. 
Throughout the paper, we adopt a cosmology with $H_0=70$ km s$^{-1}$ Mpc$^{-1}$, $\Omega_{\rm{M}}=0.3$, and $\Omega_{\Lambda}=0.7$, and a \citet{chab03} IMF. 
All magnitudes are given in the AB system.

\section{Observations and Data Reduction} \label{sec:data}

\begin{table}
\tbl{ALMA observations.}{
\begin{tabular}{ccccc}
\hline
Date       & Tuning & Sub-region& $N_{\rm ant}$ & Baseline (max) \\
           &        &           &               & (m) \\
\hline
2016-09-02 & 2      & NW        & 39, 45        & 1808.012, 2732.660 \\
2016-09-03 & 2      & NE        & 41            & 1770.782 \\
2016-09-06 & 2      & NE        & 39            & 2483.450 \\
2016-09-07 & 1      & N         & 39            & 2483.450 \\
2016-09-08 & 2      & SW        & 39            & 2483.450 \\
2016-09-12 & 2      & SE        & 38            & 3143.756 \\
2016-09-14 & 2      & SE        & 38            & 3247.644 \\
2016-09-18 & 1, 2   & NW, W     & 38            & 2483.451 \\
2016-09-19 & 2      & W         & 40            & 3143.756 \\
2016-09-20 & 1, 2   & E         & 39            & 3143.756 \\
2016-09-21 & 1, 2   & E, SW, S  & 39            & 3143.756 \\
2016-09-22 & 1, 2   & SW, S     & 39            & 3143.756 \\
2016-09-24 & 2      & N, C      & 39            & 3143.756 \\
2016-09-25 & 1, 2   & C, NE     & 39            & 3143.756 \\
2016-09-26 & 1      & NE, C     & 40            & 3247.644 \\
2016-09-27 & 1      & W, C      & 43            & 3247.644 \\
2016-09-28 & 1      & W, S, SE  & 40            & 3143.756 \\
2016-09-29 & 1      & SE        & 39            & 3247.644 \\
\hline
\end{tabular}}\label{tab:observations}
\end{table}

\begin{table}[t]
\tbl{Center frequencies of spectral windows used in the surveys of ASAGAO, HUDF \citep{dunl17}, and GOODS-ALMA \citep{fran18}.}{
\begin{tabular}{ccccc}
\hline
spw ID & \multicolumn{2}{c}{ASAGAO} & HUDF & GOODS-ALMA \\
       & tuning 1 & tuning 2 &       &  \\
       & (GHz)  & (GHz)  & (GHz) & (GHz) \\
\hline
0      & 254.12 & 245.12 & 212.2 & 255.9 \\
1      & 256.00 & 247.00 & 214.2 & 257.9 \\
2      & 269.12 & 260.12 & 228.2 & 271.9 \\
3      & 271.00 & 262.00 & 230.2 & 273.9 \\
\hline
\end{tabular}}\label{tab:tunings}
\end{table}

\begin{figure}
\begin{center}
\includegraphics[width=\linewidth]{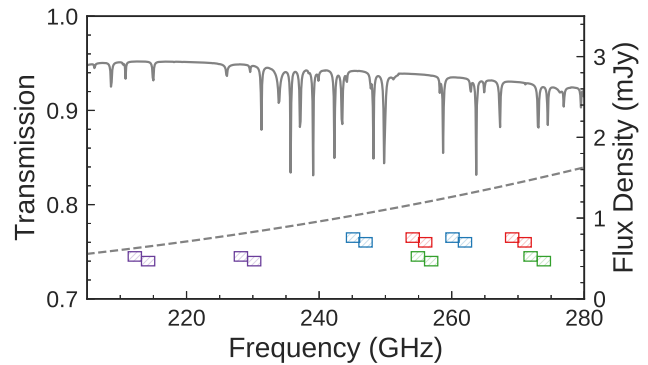}
\end{center}
\caption{
Frequency setups of ASAGAO tuning 1 (red), tuning 2 (blue), HUDF (purple), and GOODS-ALMA (green). 
Solid line represents the atmospheric transmission at the ALMA site for a precipitable water vapor of 1 mm calculated using the Atmospheric Transmission at Microwaves code (ATM; \cite{pard01})\footnotemark (left axis). 
The dashed line shows the modified black body spectrum with a dust emissivity index of $\beta = 1.5$, a dust temperature of 35~K, and $z = 2$, scaled to a flux density at 243~GHz of 1 mJy (right axis). 
}
\label{fig:tunings}
\end{figure}

\footnotetext{https://almascience.eso.org/about-alma/atmosphere-model}

\begin{figure*}
\begin{center}
\includegraphics[width=\linewidth]{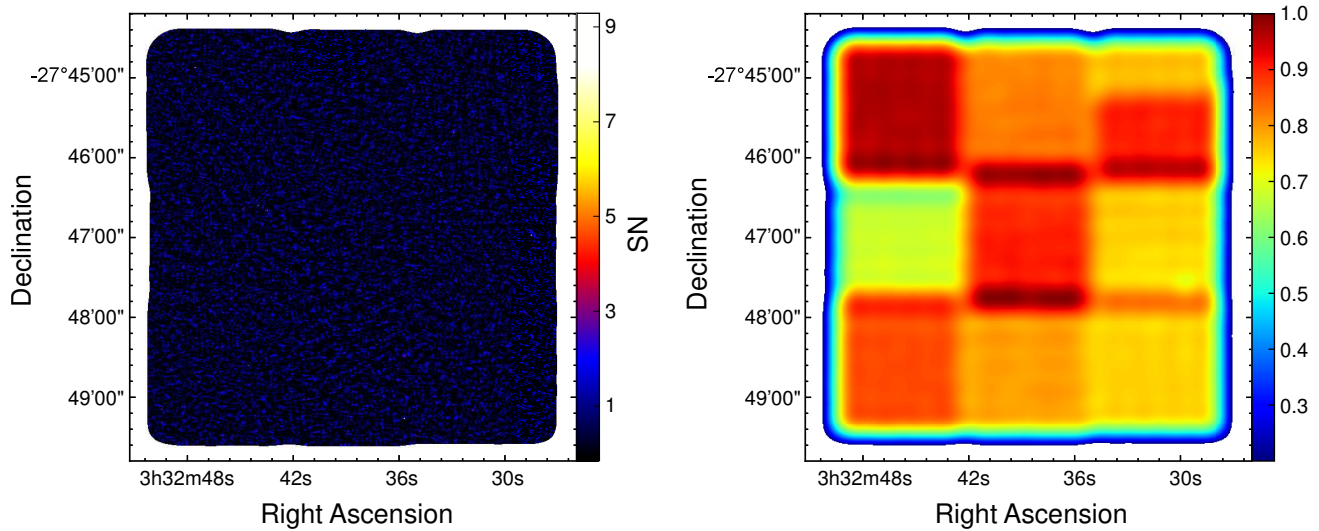}
\end{center}
\caption{
Signal-to-noise ratio map with a 250~k$\lambda$ taper (left) 
and the primary beam coverage map (right) based on the original ASAGAO data. 
}
\label{fig:asagao_map}
\end{figure*}


\subsection{Observations} \label{subsec:observations}

ALMA band 6 observations of the GOODS-S field were conducted in September 02--29, 2016 for the Cycle 3 program (Project code: 2015.1.00098.S, PI: K. Kohno) as summarized in table~\ref{tab:observations}. 
The $\sim$$5' \times 5'$ survey area centered at (R.A., Dec.) = (\timeform{03h31m38.601s}, \timeform{-27D46'59.830''}) consists of 9 tiles (figure~\ref{fig:region}) and each tile was covered by $\sim$90-pointing mosaic observations with Nyquist sampling. 
Two frequency tunings were adopted to cover a wider frequency range, providing a larger survey volume for searching serendipitous line emitting galaxies. 
The center frequencies of the tunings are 262.56~GHz (1.14~mm) and 253.56~GHz (1.18~mm), which were selected to avoid strong atmospheric absorption lines (table~\ref{tab:tunings} and figure~\ref{fig:tunings}). 
The correlator was used in the time domain mode (TDM). 
Four basebands were used for each tuning, and a spectral window (spw) was placed for each baseband with a bandwidth of 2000 MHz (15.625~MHz $\times$ 128 channels), providing a total nominal bandwidth of 16 GHz (effective bandwidth of 15 GHz) centered at 258.6~GHz (1.16~mm). 
The observations were done in 37 execution blocks in the C40-6 array configuration (maximum recoverable scale of $\theta_{\rm MRS} \approx 1\farcs2$) with a minimum baseline length of 15.065~m and a maximum baseline length ranging from 1770 m to 3247 m. 
The number of available antenna was 38--45. 
The total observing time is 45 hours, and the on-source integration time is 29 hours. 
The bandpass was calibrated with quasars J0522$-$3627, J0238+1636, and J0334$-$4008, and the phase was calibrated with J0348$-$2749. 
J0334$-$4008 and J2357$-$5311 were observed as flux calibrators.

\subsection{Data Reduction} \label{subsec:reduction}

To reduce the data volume for easier handling in continuum imaging, we average the data in frequency and time directions with 32 channels ($\Delta \nu = 0.5$~GHz) and 10.08~sec, respectively. 
The effect of bandwidth smearing on the peak flux density of a source caused by the channel averaging is less than 1\% even at the edge of the primary beam \citep{cond98}. 
We also confirm that the effect of the time averaging on the flux density is negligible based on the imaging of the bandpass calibrator.

\begin{figure}
\begin{center}
\includegraphics[width=\linewidth]{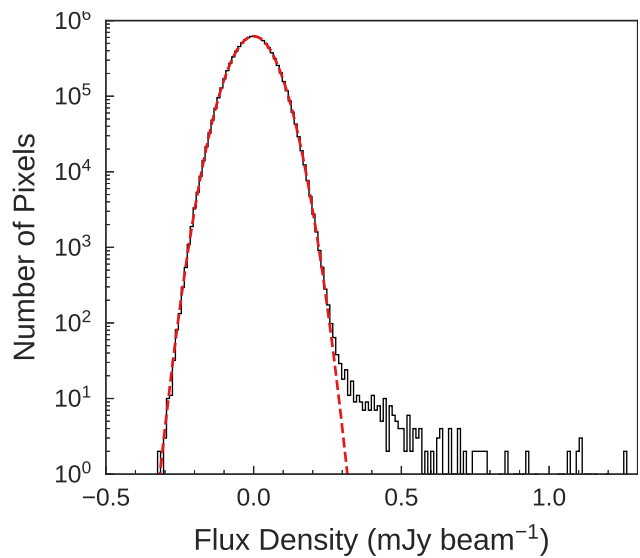}
\end{center}
\caption{
Distribution of flux density of the signal map based on the original ASAGAO data (uncorrected for primary beam attenuation). 
The dashed curve shows the result of a Gaussian fit ($1\sigma = 61$~$\mu$Jy~beam$^{-1}$). 
}
\label{fig:pix_hist}
\end{figure}

The data were reduced with Common Astronomy Software Applications (CASA; \cite{mcmu07}). 
Data calibration was done with the ALMA Science Pipeline Software of CASA version 4.7.2. 
The maps were processed by the task {\verb tclean } of CASA version 5.1.1 with natural weighting, a cell size of 0.1 arcsec, a gridding option of standard, the spectral definition mode of multi-frequency synthesis, the number of Taylor coefficients in the spectral model of 2 for a spectrum with a slope, and a primary beam limit of 0.2 (default value). 
Clean boxes are placed when a component with a peak signal-to-noise ratio (SN) above 5 is identified, and {\verb CLEAN }ed down to a $2\sigma$ level. 
Because the observations were done with a higher angular resolution ($\sim$$0.2''$) than requested because of the restriction of array configuration, we adopt a $uv$-taper of 250~k$\lambda$ to weight extended components, which gives a synthesized beam size of $0\farcs51 \times 0\farcs45$. 
The signal-to-noise ratio map and the primary beam coverage map are shown in figure~\ref{fig:asagao_map}. 
In this study, we use the region where the primary beam coverage is larger than or equal to 0.2 in the map, which is a 26-arcmin$^2$ area. 
A sensitivity map was created by using the {\sc BANE} program \citep{hanc12}, which performs $3\sigma$ clipping in the signal map and calculate the standard deviation on a sparse grid of pixels and then interpolate to make a noise image. 
Figure~\ref{fig:pix_hist} shows the histograms of flux density of the signal map (before primary beam correction).
The pixel-flux distribution is well explained by a Gaussian curve, and a Gaussian fit gives $1\sigma$ of 61~$\mu$Jy~beam$^{-1}$. 
The excess from the fitted Gaussian at $\gtrsim$0.3~mJy indicates the contribution from real sources.

\subsection{ALMA Archival Data}

In addition to our data, we also use the ALMA archival data of 1-mm (band 6) surveys in the GOODS-S field of HUDF \citep{dunl17} and GOODS-ALMA \citep{fran18}. 
We do not use the data set of ASPECS, where the synthesized beam size ($1\farcs68 \times 0\farcs92$) is largely different from those of the others ($\lesssim 0.5''$).

The ALMA survey of HUDF by \citet{dunl17} covered a 4.5 arcmin$^2$ area at 1.3~mm during Cycle 1 and 2 (Project code: 2012.1.00173.S, PI: J. Dunlop). 
The correlator was configured with four spectral windows with a 2000 MHz bandwidth (15.625 MHz $\times$ 128 channels). 
The synthesized beam with natural weighting is $0\farcs59 \times 0\farcs50$. 
An $uv$-tapering of $\simeq$$220 \times 180$~k$\lambda$ they adopted gives a final synthesized beam of $0\farcs71 \times 0\farcs67$ and a noise level of 34~$\mu$Jy~beam$^{-1}$.

A wider area of 69 arcmin$^2$ ($\sim$$10' \times 7'$) was observed in the GOODS-ALMA survey \citep{fran18} at 1.13~mm during Cycle 3 (Project code: 2015.1.00543.S, PI: D. Elbaz). 
The survey consists of six sub-mosaics, encompassing the survey fields of ASAGAO, HUDF, and ASPECS. 
The correlator was set to have four spectral windows with 15.625 MHz $\times$ 128 channels. 
The synthesized beam with natural weighting is $\sim$0\farcs20--0\farcs29 depending on the sub-regions. 
The rms noise level is $\sim$180~$\mu$Jy~beam$^{-1}$ and $\sim$110~$\mu$Jy~beam$^{-1}$ for the tapered map with a synthesized beam of 0\farcs6 and for the untapered map, respectively.

\begin{figure}
\begin{center}
\includegraphics[width=\linewidth]{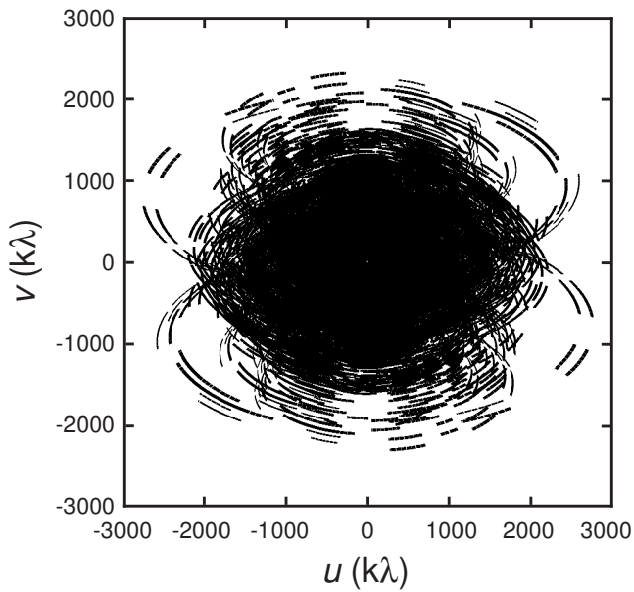}
\end{center}
\caption{
$uv$-plane coverage of the combined data. 
}
\label{fig:uv}
\end{figure}

\begin{figure*}
\begin{center}
\includegraphics[width=.8\linewidth]{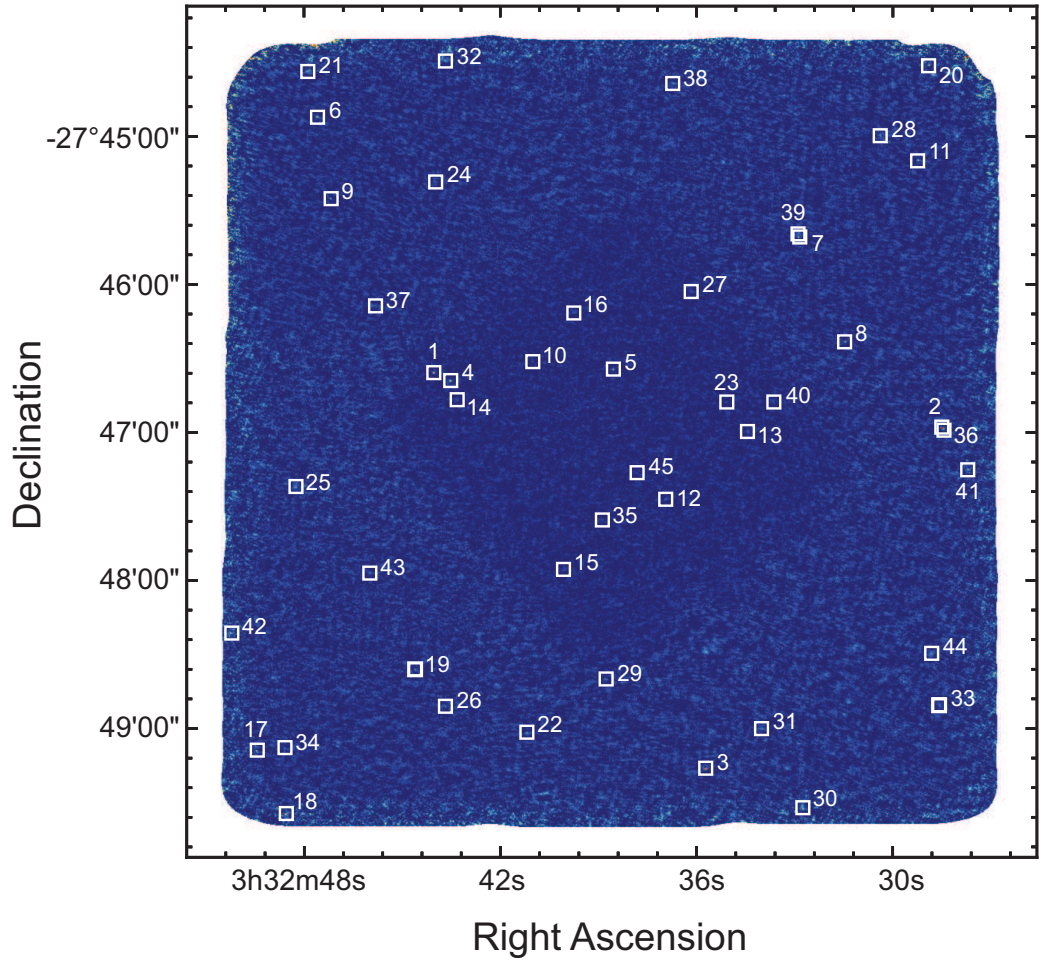}\\
\end{center}
\caption{
The combined signal map (ASAGAO + HUDF + GOODS-ALMA) with a 250 k$\lambda$ taper (corrected for primary beam attenuation). 
The squares represent the detected sources. 
}
\label{fig:combined_sigmap}
\end{figure*}

\begin{figure*}
\begin{center}
\includegraphics[width=\linewidth]{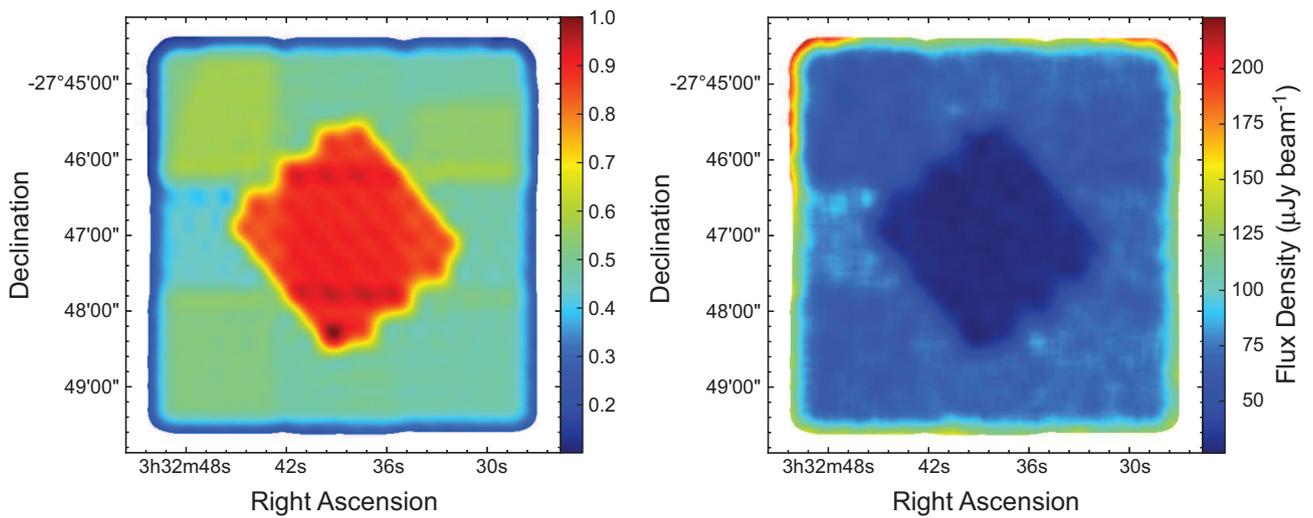}
\end{center}
\caption{
The primary beam coverage map (left) and the rms noise map (right) for the combined data. 
The rms noise map is corrected for primary beam attenuation. 
}
\label{fig:combined_covmap}
\end{figure*}

\begin{figure}
\begin{center}
\includegraphics[width=\linewidth]{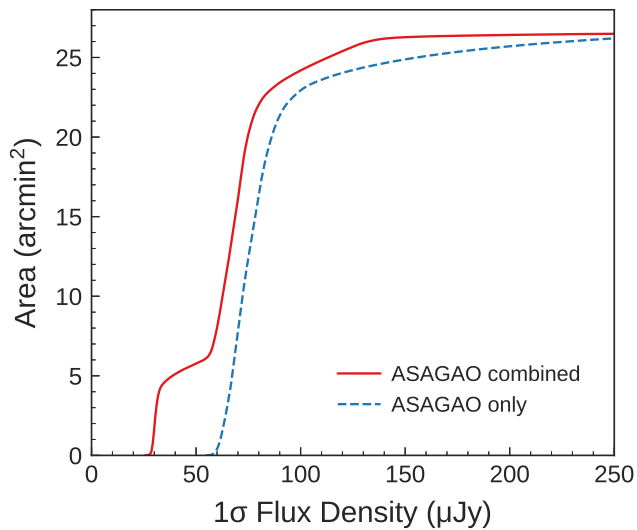}
\end{center}
\caption{
Cumulative area of the combined map as a function of rms noise level (corrected for primary beam attenuation) for the ASAGAO only data (dashed) and the combined data (solid). 
}
\label{fig:effective_area}
\end{figure}

\begin{figure}
\begin{center}
\includegraphics[width=\linewidth]{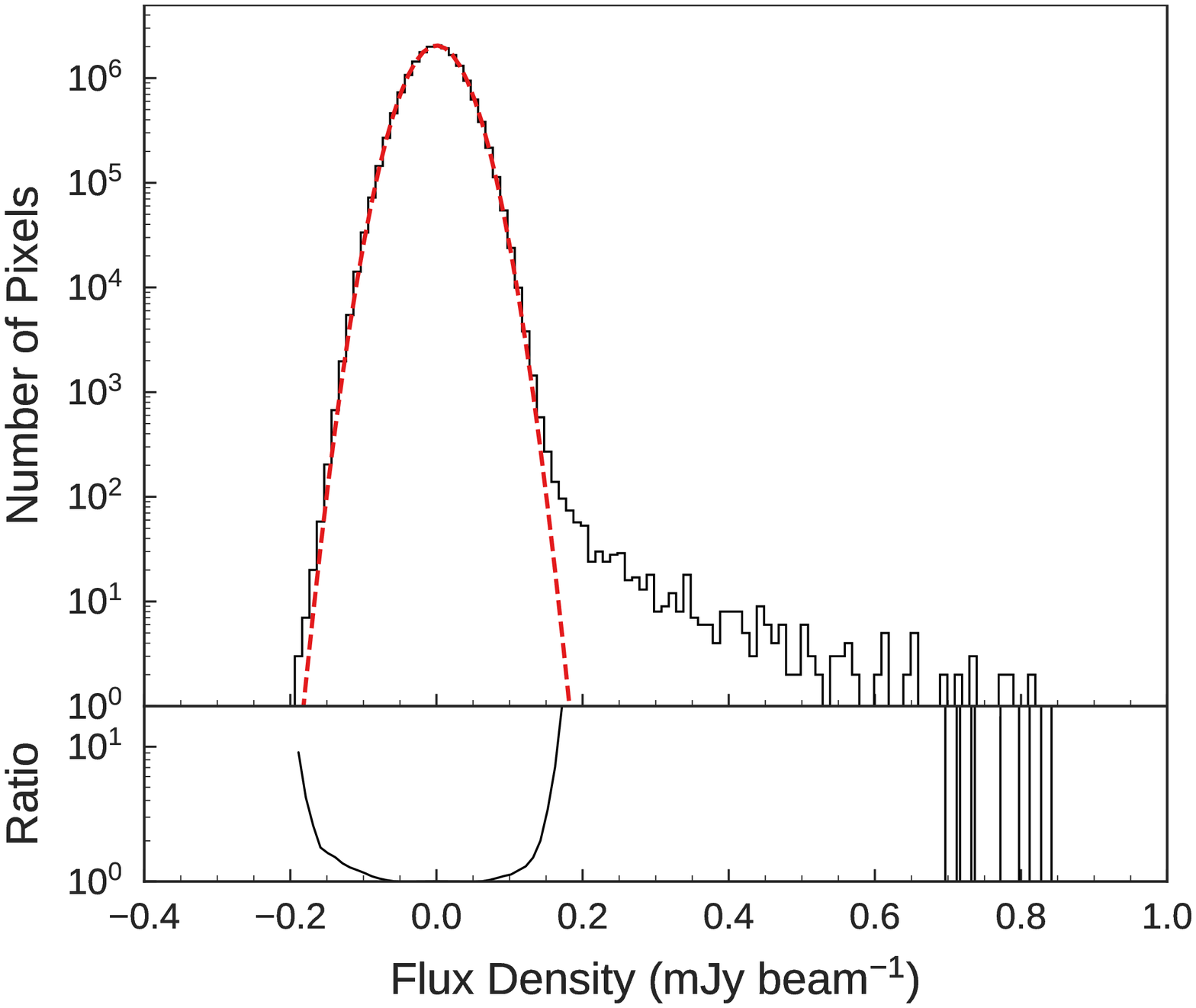}
\end{center}
\caption{
\textit{Top}: Distribution of flux density of the signal map based on the combined data (uncorrected for primary beam attenuation). 
The dashed curve shows the result of a Gaussian fit ($1\sigma = 34$~$\mu$Jy~beam$^{-1}$). 
\textit{Bottom}: Ratio between the flux density distribution and the result of a Gaussian fit. 
}
\label{fig:pix_hist_combined}
\end{figure}

\subsection{Combined Map} \label{subsec:combined}

The archival data sets of HUDF \citep{dunl17} and GOODS-ALMA \citep{fran18} are combined to the original ASAGAO data to make a deeper map with the total effective frequency coverage of $\sim$27 GHz (table~\ref{fig:tunings} and figure~\ref{tab:tunings}). 
Before combining the data sets, we relabel the coordinates of Cycle 1 and 2 data from J2000.0 to the International Celestial Reference System (ICRS) by using a CASA script offered by the ALMA project, because the position reference frame in ALMA $uv$ data and images is given as J2000 before Cycle 3 and as ICRS from Cycle 3. 
The $uv$ data sets are averaged in frequency and time directions (32 channels and 10.08~sec) in the same manner as the original ASAGAO data. 
Figure~\ref{fig:uv} shows the $uv$-plane coverage of the combined data. 
The combined map was produced with CASA with the same parameters adopted in Sec~\ref{subsec:reduction}. 
The representative frequency of the map is 243.047 GHz (1.23 mm). 
We adopt an $uv$-taper of 250~k$\lambda$ to weight extended components, which gives a final synthesized beam size of $0\farcs59 \times 0\farcs53$. 
Maps without $uv$-taper (synthesized beam size of $0\farcs30 \times 0\farcs24$) and with a $uv$-taper of 160~k$\lambda$ ($0\farcs83 \times 0\farcs72$) were also created to see whether detected sources are spatially resolved. 
The signal map, the coverage map, and the rms noise map (corrected for primary beam attenuation) with a 250~k$\lambda$ taper are shown in figure~\ref{fig:combined_sigmap} and \ref{fig:combined_covmap}. 
We use the same region as adopted in the original ASAGAO map (Sec.~\ref{subsec:reduction}). 
The map has two layers, the central deeper area (the deepest region has $1\sigma \sim 26$~$\mu$Jy~beam$^{-1}$) and the rest, as can be seen in figure~\ref{fig:combined_covmap} and figure~\ref{fig:effective_area} of the cumulative area as a function of rms noise level.

Figure~\ref{fig:pix_hist_combined} shows the histogram of flux density of the signal map (before primary beam correction).
The dashed curve represents the result of a Gaussian fit, which gives $1\sigma$ of 34~$\mu$Jy~beam$^{-1}$. 
The presence of real sources in the map makes excess of positive pixels. 
This fit also deviates from the distribution of pixel values at high negative flux densities, which can be explained by the non-uniform noise distribution of the entire map.

\section{Source Catalog} \label{sec:source}

\subsection{Source Detection} \label{subsec:detection}

Source detection is conducted on the signal map before correcting for the primary beam attenuation. 
We adopt the source-finding algorithm called {\sc Aegean} \citep{hanc12, hanc18}, which achieves high reliability and completeness performance for radio maps. 
The background and noise estimation are done with the {\sc BANE} package in the same manner as described in Sec.~\ref{subsec:reduction}. 
We find 25 (45) sources with a peak SN of $\ge$5$\sigma$ ($\ge$4.5$\sigma$). 
The detected sources are fitted with a 2D elliptical Gaussian to estimate the source size and integrated flux density. 
The integrated flux density ($S_{\rm int}$) is calculated as
\begin{eqnarray}
S_{\rm int} = S_{\rm peak} \frac{a b}{\theta_{\rm maj}\theta_{\rm min}}, 
\end{eqnarray}
where $S_{\rm peak}$ is the peak flux density, $a/b$ are the fitted major/minor axes, and $\theta_{\rm maj}/\theta_{\rm min}$ are the synthesized beam major/minor axes. 
We adopt $S_{\rm int}$ as the source flux density. 
When $S_{\rm int} < S_{\rm peak}$, we adopt $S_{\rm peak}$, since it is possible that the source fitting failed due to the low SN.

\begin{figure*}
\begin{center}
\includegraphics[width=.8\linewidth]{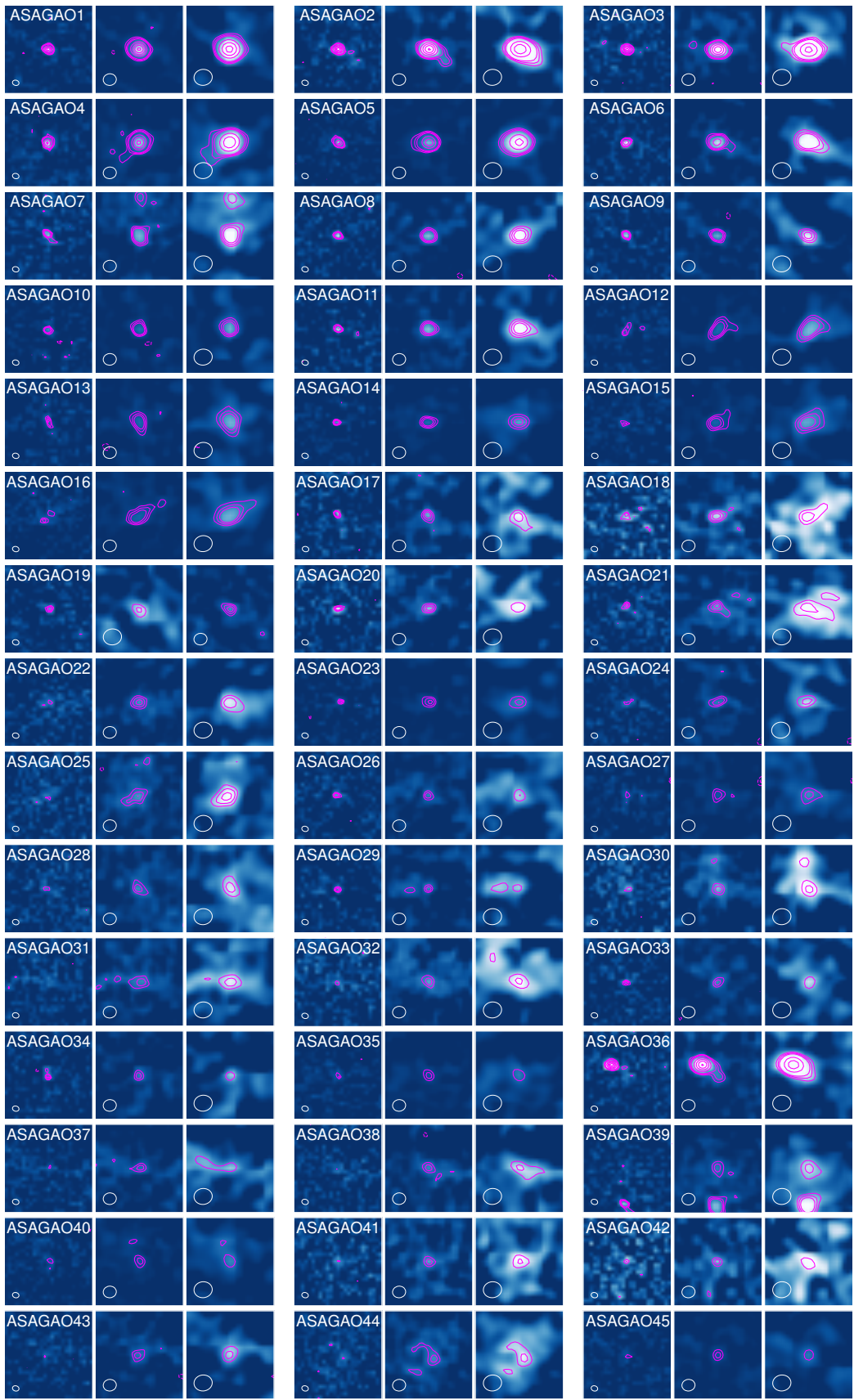}
\end{center}
\caption{
Postage-stamp images of the ASAGAO $\ge$4.5$\sigma$ sources with no taper (left), a 250-k$\lambda$ taper (middle), and a 160-k$\lambda$ taper (right). 
The image size is $4'' \times 4''$. 
Contours are 3$\sigma$, 4$\sigma$, 5$\sigma$, and 5$\sigma$ steps subsequently (negative contours are shown as dashed lines). 
The synthesized beam size is shown in the lower left corners of each panel. 
\label{fig:source}}
\end{figure*}

The source catalog for the 4.5$\sigma$ sources extracted in the combined signal map with a 250 k$\lambda$ taper is presented in table~\ref{tab:source}. 
Hereafter we refer to these sources as ASAGAO sources, and adopt the integrated flux densities measured in the 250 k$\lambda$ tapered map. 
The range of continuum flux densities is 0.16--2 mJy (after correcting for primary beam attenuation). 
The integrated flux densities in the untapered map ($S_{\rm int}^{\rm untaper}$) and in the map with a 160 k$\lambda$ taper ($S_{\rm int}^{\rm 160k\lambda}$) measured in the same manner as in the 250 k$\lambda$ tapered map are also shown. 
When a source is not detected with a peak SN $>$ 3 in these maps, the flux density is not listed in the source catalog. 
ASAGAO ID31, 36, and 37 are not detected in the untapered map with a peak SN $>$ 3. 
This can be due to the lack of sensitivity for spatially extended structures or clumpy structures and multiple peaks as can be seen in the postage-stamp images in figure~\ref{fig:source}, each having a peak SN less than 3. 
The median ratio between integrated flux and peak flux is $S_{\rm int}/S_{\rm peak} = 1.3 \pm 0.8$. 
The median ratio of integrated flux between 250~k$\lambda$-tapered map and 160~k$\lambda$-tapered map or the untapered map is $S_{\rm int}^{\rm 250k\lambda}/S_{\rm int}^{\rm 160k\lambda} = 0.86 \pm 0.24$, and $S_{\rm int}^{\rm 250k\lambda}/S_{\rm int}^{\rm untaper} = 1.3 \pm 3.0$. 
These suggest that sources are resolved by the synthesized beam in the 250 k$\lambda$-tapered and the untapered maps.

\begin{figure}
\begin{center}
\includegraphics[width=\linewidth]{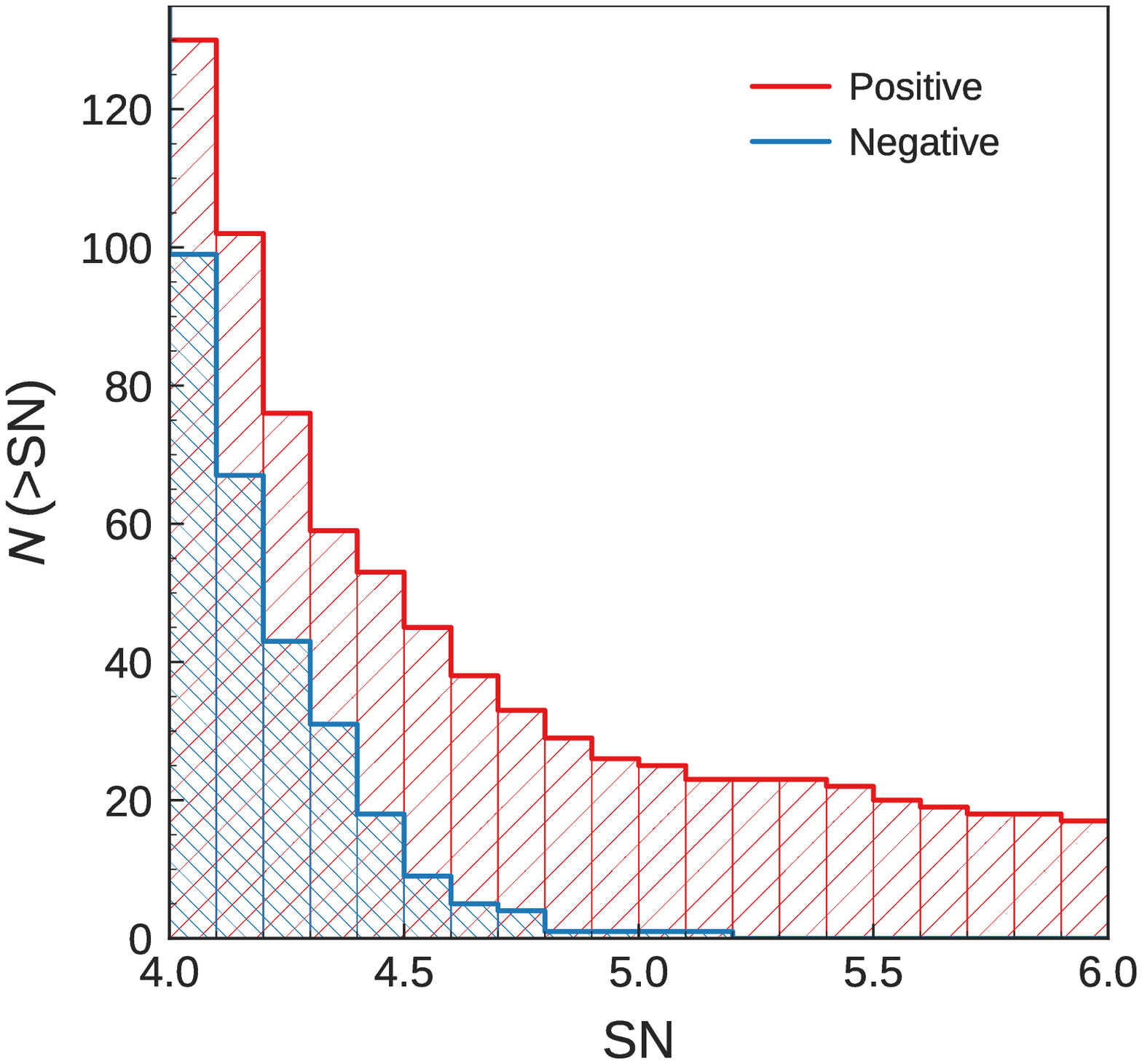}
\end{center}
\caption{
Cumulative number of positive and negative peaks as a function of peak SN threshold. 
}
\label{fig:fdr}
\end{figure}

In order to estimate the degree of contamination by spurious sources, we count the number of negative peaks as a function of SN threshold (figure~\ref{fig:fdr}). 
The number of independent beams in the map is $2.7 \times 10^5$, and the expected number of $\ge$$4.5\sigma$ sources in a Gaussian statistics is $\sim$1. 
However, it is reported that this estimation underestimates the negative peaks in previous studies based on ALMA images \citep{dunl17, vio16, vio17}. 
The actual number of negative peaks in the combined map is 1 at $\ge$5$\sigma$ and 8 at 4.5--5$\sigma$.

The small number of negative peaks at $\ge$5$\sigma$ suggests the robustness of the 5$\sigma$ sources. 
Actually, 22 out of the 25 5$\sigma$ sources (88\%) have counterparts at optical, {\sl Spitzer}/IRAC, radio, or ALMA 850~$\mu$m \citep{cowi18} (see \cite{yama18} for multi-wavelength identifications of ASAGAO sources).

\begin{table*}[t]
\tbl{Source catalog of $\ge$5$\sigma$ sources (ID1--25) and 4.5--5$\sigma$ sources (ID26--45)}{
\footnotesize
\begin{tabular}{ccccccccc}
\hline
ID&R.A.&Dec.&SN&$S_{\rm peak}$&$S_{\rm int}$&$S_{\rm int}^{\rm untaper}$&$S_{\rm int}^{\rm 160k\lambda}$&Note\\
ASAGAO&(J2000)&(J2000)&  &($\mu$Jy)&($\mu$Jy)&($\mu$Jy)&($\mu$Jy)&  \\
(1)&(2)&(3)&(4)&(5)&(6)&(7)&(8)&(9)\\
\hline
 1 & 03:32:44.03 & $-$27:46:35.97 & 26.0 & $ 839 \pm  32$ & $ 990 \pm  36$ & $ 877 \pm  32$ & $1023 \pm  44$ & UDF1, AGS6, U3 \\
 2 & 03:32:28.51 & $-$27:46:58.36 & 25.6 & $1851 \pm  72$ & $1983 \pm  75$ & $1996 \pm  57$ & $2251 \pm 112$ & AGS1, U1 \\
 3 & 03:32:35.72 & $-$27:49:16.27 & 24.0 & $1656 \pm  69$ & $1758 \pm  70$ & $1816 \pm  54$ & $1709 \pm 100$ & AGS3, U2 \\
 4 & 03:32:43.53 & $-$27:46:39.25 & 21.0 & $ 658 \pm  31$ & $ 914 \pm  41$ & $ 761 \pm  40$ & $1019 \pm  50$ & UDF2, AGS18, U6 \\
 5 & 03:32:38.55 & $-$27:46:34.61 & 18.1 & $ 554 \pm  31$ & $ 745 \pm  39$ & $ 634 \pm  35$ & $ 791 \pm  45$ & UDF3, ASPECS/C1, AGS12, U8 \\
 6 & 03:32:47.59 & $-$27:44:52.43 & 12.4 & $ 768 \pm  62$ & $ 954 \pm  74$ & $ 735 \pm  43$ & $1161 \pm 123$ & U4  \\
 7 & 03:32:32.90 & $-$27:45:41.07 &  8.8 & $ 546 \pm  63$ & $ 829 \pm  86$ & $ 593 \pm  65$ & $ 835 \pm 104$ & U5  \\
 8 & 03:32:31.48 & $-$27:46:23.50 &  8.7 & $ 576 \pm  66$ & $ 650 \pm  72$ & $ 618 \pm  57$ & $ 705 \pm 103$ & AGS13, U12 \\
 9 & 03:32:47.18 & $-$27:45:25.48 &  8.6 & $ 495 \pm  57$ & $ 488 \pm  55$ & $ 945 \pm 106$ & $ 406 \pm  69$ & \\
10 & 03:32:41.02 & $-$27:46:31.59 &  8.6 & $ 255 \pm  30$ & $ 278 \pm  31$ & $ 350 \pm  39$ & $ 246 \pm  32$ & UDF4 \\
11 & 03:32:29.25 & $-$27:45:09.96 &  8.5 & $ 580 \pm  68$ & $ 678 \pm  78$ & $ 587 \pm  51$ & $ 855 \pm 124$ & \\
12 & 03:32:36.96 & $-$27:47:27.14 &  7.4 & $ 227 \pm  31$ & $ 408 \pm  49$ & $ 190 \pm  35$ & $ 484 \pm  61$ & UDF5 \\
13 & 03:32:34.44 & $-$27:46:59.86 &  7.2 & $ 224 \pm  31$ & $ 436 \pm  53$ & $ 227 \pm  34$ & $ 503 \pm  65$ & UDF6 \\
14 & 03:32:43.33 & $-$27:46:46.96 &  7.2 & $ 229 \pm  32$ & $ 259 \pm  35$ & $ 224 \pm  26$ & $ 281 \pm  48$ & UDF7, U7 \\
15 & 03:32:40.07 & $-$27:47:55.72 &  6.6 & $ 197 \pm  30$ & $ 458 \pm  64$ & $ 166 \pm  32$ & $ 490 \pm  69$ & UDF11 \\
16 & 03:32:39.75 & $-$27:46:11.67 &  6.5 & $ 192 \pm  29$ & $ 539 \pm  65$ & $ 106 \pm  22$ & $ 640 \pm  76$ & UDF8, ASPECS/C2 \\
17 & 03:32:49.45 & $-$27:49:09.00 &  6.1 & $ 516 \pm  83$ & $ 564 \pm  90$ & $ 485 \pm  55$ & $1286 \pm 289$ & U11 \\
18 & 03:32:48.57 & $-$27:49:34.62 &  5.8 & $ 749 \pm 130$ & $1091 \pm 172$ & $ 353 \pm  94$ & $1868 \pm 370$ & \\
19 & 03:32:44.61 & $-$27:48:36.13 &  5.7 & $ 375 \pm  67$ & $ 434 \pm  73$ & $ 345 \pm  51$ & $ 431 \pm  99$ & U10 \\
20 & 03:32:28.91 & $-$27:44:31.54 &  5.6 & $ 614 \pm 109$ & $ 653 \pm 110$ & $ 637 \pm  80$ & $ 749 \pm 187$ & \\
21 & 03:32:47.90 & $-$27:44:33.96 &  5.5 & $ 499 \pm  91$ & $1011 \pm 178$ & $ 356 \pm  57$ & $3116 \pm 536$ & \\
22 & 03:32:41.20 & $-$27:49:01.75 &  5.4 & $ 371 \pm  68$ & $ 612 \pm 101$ & $ 187 \pm  43$ & $ 797 \pm 153$ & \\
23 & 03:32:35.09 & $-$27:46:47.82 &  5.4 & $ 163 \pm  30$ & $ 206 \pm  37$ & $ 135 \pm  20$ & $ 202 \pm  44$ & UDF13 \\
24 & 03:32:43.99 & $-$27:45:18.74 &  5.0 & $ 299 \pm  60$ & $ 446 \pm  82$ & $  65 \pm 120$ & $ 482 \pm 102$ & \\
25 & 03:32:48.24 & $-$27:47:22.14 &  5.0 & $ 385 \pm  77$ & $ 858 \pm 223$ & $ 186 \pm  46$ & $1168 \pm 186$ & \\
\hline
26 & 03:32:43.68 & $-$27:48:51.12 &  4.9 & $ 314 \pm  64$ & $ 254 \pm  52$ & $ 286 \pm  37$ & $ 364 \pm  91$ & \\
27 & 03:32:36.17 & $-$27:46:03.04 &  4.9 & $ 154 \pm  32$ & $ 226 \pm  45$ & $ 170 \pm  43$ & $ 374 \pm  80$ & \\
28 & 03:32:30.41 & $-$27:44:59.97 &  4.9 & $ 348 \pm  72$ & $ 716 \pm 154$ & $ 124 \pm  34$ & $ 888 \pm 184$ & \\
29 & 03:32:38.74 & $-$27:48:40.12 &  4.8 & $ 348 \pm  71$ & $ 227 \pm  46$ & $ 551 \pm  84$ & $ 677 \pm 185$ & \\
30 & 03:32:32.76 & $-$27:49:32.41 &  4.7 & $ 578 \pm 122$ & $ 886 \pm 180$ & $ 452 \pm 123$ & $1157 \pm 253$ & \\
31 & 03:32:34.02 & $-$27:49:00.11 &  4.7 & $ 339 \pm  72$ & $ 846 \pm 158$ & --             & $ 881 \pm 173$ & \\
32 & 03:32:43.68 & $-$27:44:29.66 &  4.7 & $ 461 \pm  98$ & $ 769 \pm 164$ & $ 299 \pm  56$ & $1140 \pm 249$ & \\
33 & 03:32:28.59 & $-$27:48:50.57 &  4.7 & $ 347 \pm  74$ & $ 366 \pm  79$ & $ 283 \pm  42$ & $ 425 \pm 108$ & \\
34 & 03:32:48.60 & $-$27:49:07.95 &  4.6 & $ 298 \pm  65$ & $ 313 \pm  67$ & $ 303 \pm  52$ & $ 328 \pm  80$ & \\
35 & 03:32:38.89 & $-$27:47:35.50 &  4.6 & $ 140 \pm  30$ & $ 180 \pm  39$ & $  74 \pm  18$ & $ 169 \pm  44$ & \\
36 & 03:32:28.46 & $-$27:46:58.83 &  4.6 & $ 333 \pm  73$ & $ 635 \pm 134$ & --             & --             & \\
37 & 03:32:45.83 & $-$27:46:08.86 &  4.6 & $ 270 \pm  59$ & $ 362 \pm  76$ & --             & $1879 \pm 405$ & \\
38 & 03:32:36.74 & $-$27:44:38.73 &  4.6 & $ 334 \pm  73$ & $ 441 \pm  91$ & $ 175 \pm  50$ & $1496 \pm 309$ & \\
39 & 03:32:32.90 & $-$27:45:39.37 &  4.6 & $ 286 \pm  63$ & $ 529 \pm 107$ & $ 148 \pm  37$ & $ 833 \pm 194$ & \\
40 & 03:32:33.65 & $-$27:46:47.94 &  4.6 & $ 149 \pm  33$ & $ 198 \pm  42$ & $ 304 \pm  66$ & $ 202 \pm  51$ & \\
41 & 03:32:27.72 & $-$27:47:15.17 &  4.6 & $ 459 \pm  99$ & $ 629 \pm 133$ & $ 229 \pm  60$ & $ 805 \pm 188$ & \\
42 & 03:32:50.25 & $-$27:48:21.16 &  4.6 & $ 588 \pm 129$ & $ 621 \pm 137$ & $ 596 \pm 137$ & $ 885 \pm 224$ & \\
43 & 03:32:45.99 & $-$27:47:57.18 &  4.6 & $ 283 \pm  62$ & $ 495 \pm 106$ & $ 149 \pm  42$ & $ 589 \pm 128$ & \\
44 & 03:32:28.84 & $-$27:48:29.72 &  4.5 & $ 350 \pm  77$ & $2051 \pm 447$ & $ 107 \pm  25$ & $1679 \pm 362$ & \\
45 & 03:32:37.83 & $-$27:47:16.49 &  4.5 & $ 131 \pm  29$ & $ 157 \pm  33$ & $ 117 \pm  25$ & $ 128 \pm  33$ & \\
\hline
\end{tabular}}\label{tab:source}
\begin{tabnote}
Notes. - 
(1) ASAGAO ID. 
(2) Right ascension. 
(3) Declination. 
(4) Peak signal-to-noise ratio. 
(5) Peak flux density (corrected for primary beam attenuation). 
(6) Integrated flux density (corrected for primary beam attenuation). 
(7) Integrated flux density (corrected for primary beam attenuation) measured in the untapered map when the peak SN is above 3. 
(8) Integrated flux density (corrected for primary beam attenuation) measured in the 160-k$\lambda$ tapered map when the peak SN is above 3. 
(9) Notes on source IDs of \citet{dunl17} (UDF), \citet{arav16} (ASPECS), \citet{fran18} (AGS), and \citet{ueda18} (U). 
\end{tabnote}
\end{table*}

\subsection{Astrometry} \label{subsec:astrometry}

Calibration for astrometry is performed by interpolating the phase information of the phase calibrators over the target fields. 
The astrometric accuracy of a source depends on statistical errors determined by the source SN and systematic errors such as the atmospheric phase stability, the proximity of an astrometric calibrator, and baseline errors. 
The minimum obtainable astrometric accuracy with no systematic errors is determined by a source SN, observing frequency, and maximum baseline length, which gives $\sim$$0.15''$ for a 5$\sigma$ source with the observing frequency of 243.047 GHz and the maximum baseline of 3.2 km (see ALMA Technical Handbook).

\begin{figure}
\begin{center}
\includegraphics[width=\linewidth]{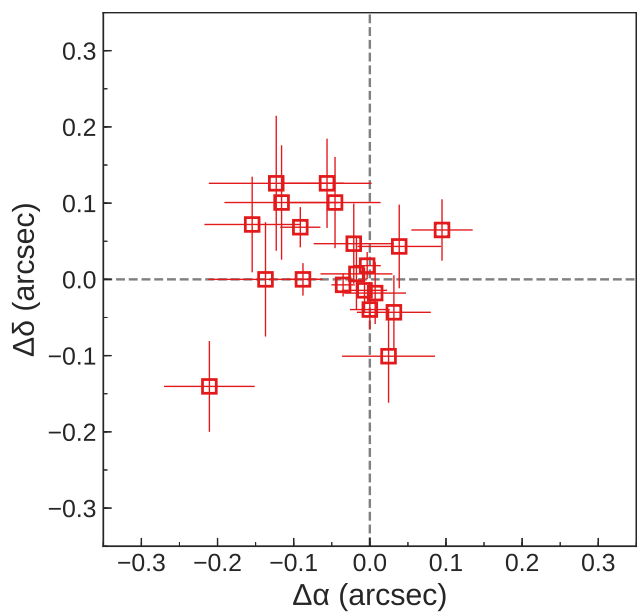}
\end{center}
\caption{
Positional offsets of the 5$\sigma$ ASAGAO sources from the VLA 5-cm radio sources (\cite{rujo16}, Rujopakarn et al. in prep.). 
The errors are the square root of the sum of the squares of expected 1$\sigma$ positional uncertainties of the ASAGAO and VLA sources. 
\label{fig:offset}}
\end{figure}

To confirm the astrometry of ASAGAO sources, 
the positions of the 5$\sigma$ sources are cross-matched with sources detected in the VLA 5-cm survey (\cite{rujo16}, Rujopakarn et al. in prep.). 
The radio sources are more suitable for evaluating the astrometry of the ALMA sources compared to optical sources because 
(i) the angular resolution and positional accuracy are comparable to those of the ALMA observations, 
and (ii) the positions of submm/mm emission and optical emission, which typically trace dust obscured and unobscured parts, respectively, do not necessarily coincide within a galaxy, and radio observations can trace dust obscured parts. 
The radio counterparts are found for 20 out of the 25 ASAGAO 5$\sigma$ sources within a 0\farcs5 search radius, and the positional offset between them is plotted in figure~\ref{fig:offset}. 
The median offset is ($\Delta \alpha$, $\Delta \delta$) $= (+0\farcs03 \pm 0\farcs08, -0\farcs01 \pm 0\farcs06)$, which is within the expected positional uncertainty between the ALMA and the radio sources of $\sim$0\farcs1 as the square-root of sum of squares of both uncertainties ($\Delta \alpha = \Delta \delta \simeq 0.6$ (SN)$^{-1}$ FWHM; \cite{ivis07}).

\subsection{Comparison with ALMA 1-mm Sources in GOODS-S}
We cross-matched the ASAGAO sources with the HUDF, GOODS-ALMA, and ASPECS sources (Table~\ref{tab:source}). 
\citet{dunl17} listed 16 HUDF sources, and we confirmed that all the eight sources with SN $> 4.5$ out of 16 are detected in our map. 
Two additional other sources are detected in our map, and the other 6 sources are not detected due to their lower SNs. 
Among the 20 GOODS-ALMA sources presented in \citet{fran18}, we confirmed that all of the six sources inside the ASAGAO region are detected in our map. 
A comparison of flux densities of sources common with these surveys shows that the median flux ratios are $S_{\rm 243GHz}^{\rm ASAGAO}/S_{\rm 221GHz}^{\rm HUDF} = 1.15 \pm 0.64$ and $S_{\rm 243GHz}^{\rm ASAGAO}/S_{\rm 265GHz}^{\rm GOODS-ALMA} = 0.89 \pm 0.13$, which are consistent with the flux ratios assuming a modified black body with a dust emissivity index of $\beta = 1.5$, a dust temperature of 35~K, and $z = 2$ ($S_{\rm 243GHz}/S_{\rm 221GHz} = 1.3$ and $S_{\rm 243GHz}/S_{\rm 265GHz} = 0.78$). 
The two brightest sources ($S_{\rm 1.2mm} > 0.2$ mJy) of ASPECS, which are the highest SN sources (SN $>$ 10) in their source catalog, are also detected in our map. 
The non-detection of lower SN ASPECS sources can be explained by their lower flux densities ($S_{\rm 1.2mm} < 0.15$ mJy). 
The ASAGAO 5$\sigma$ sources without counterpart in the other surveys are outside the regions of ASPECS and HUDF, and have lower flux densities than the detection limit of GOODS-ALMA.

\subsection{Comparison with AzTEC Sources} \label{subsec:aztec}

The central 270~arcmin$^2$ area of the GOODS-S field was observed with AzTEC \citep{wils08}, mounted on the Atacama Submillimeter Telescope Experiment (ASTE; \cite{ezaw04, ezaw08}) at 1.1 mm (270 GHz) \citep{scot10}. 
The beam size of AzTEC on ASTE is 30$''$ (FWHM). 
Two AzTEC sources identified in \citet{scot10} (AzTEC/GS18 and 21) are located inside the ASAGAO region, and detected as multiple sources in our 4.5$\sigma$ source catalog.

AzTEC/GS18 is detected as three ASAGAO sources (ID1, 4, and 14), and the total flux of the three sources is $S_{\rm 1.2mm} = 2.16 \pm 0.06$~mJy, which is consistent with the flux density of the AzTEC source, $S_{\rm 1.1mm} = 3.2 \pm 0.6$~mJy \citep{down12} taking into account the flux ratio between 1.2 mm and 1.1 mm of $S_{\rm 1.2mm}/S_{\rm 1.1mm} \sim 0.73$. 
\citet{yun12} studied the radio and {\sl Spitzer} counterparts of the AzTEC/GOODS-S sources. 
They found three counterpart candidates for AzTEC/GS18, two of which are detected in the ASAGAO map. 
The other is identified in the 1.3 mm source catalog of \citet{dunl17} as a 4.26$\sigma$ source (UDF9).

AzTEC/GS21 has an ASAGAO counterpart (ID6) within $15''$ from the AzTEC source position. 
Another source (ID21) is located $\sim$$16''$ away from the AzTEC source position. 
ASAGAO ID6 is identified as a radio and {\sl Spitzer} counterpart candidate of \citet{yun12}. 
The total flux of the two ALMA sources is $S_{\rm 1.2mm} = 1.97 \pm 0.19$~mJy, which is also consistent with the flux density of the AzTEC source, $S_{\rm 1.1mm} = 2.7 \pm 0.6$~mJy \citep{down12} by considering the expected flux ratio between 1.2 mm and 1.1 mm emission.

\section{Number Counts}\label{sec:counts}

Number counts are constructed by using the 45 4.5$\sigma$ sources. 
We correct for the effective area where sources are detected at SN $\ge$ 4.5, contribution of spurious sources, survey completeness, and flux boosting. 
In this section, we present the methods of estimating survey completeness and flux boosting (Sec.~\ref{subsec:completeness}), and constructing number counts (Sec.~\ref{subsec:counts}). 
Next we compare the obtained number counts with previous studies (Sec.~\ref{subsec:counts_comparison}) and estimate the contribution of the ASAGAO sources to the 1.2~mm EBL (Sec.~\ref{subsec:ebl}).

\subsection{Completeness and Flux Boosting}\label{subsec:completeness}

We calculate the completeness, which is the rate at which a source is expected to be detected in a map, to see the effect of noise fluctuations on the source detection. 
The calculation is conducted on the signal map (corrected for primary beam attenuation). 
An artificial source of an elliptical Gaussian with the synthesized beam size is injected into a position randomly selected in the map. 
In order to take into account the effect of source size, the input source is convolved with another Gaussian function. 
\citet{fran18} computed the completeness with different convolving Gaussian FWHM between 0\farcs2 and 0\farcs9, and found that the completeness is lower for a larger FWHM. 
Recent ALMA measurements of source size of SMGs ($S_{\rm 1mm} > 1$~mJy) show that source sizes (FWHM) range from $0\farcs08$ to $0\farcs8$ (e.g., \cite{ikar15, simp15, hodg16, ikar17, umeh17}). 
The median source sizes in these studies are $0\farcs20^{+0\farcs03}_{-0\farcs05}$ \citep{ikar15}, $0\farcs30 \pm 0\farcs04$ \citep{simp15}, and $0\farcs31 \pm 0\farcs03$ \citep{ikar17}. 
\citet{fuji17} find a positive correlation between the effective radius in the rest-frame FIR wavelength and FIR luminosity by using a sample of 1034 ALMA sources, suggesting that the ASAGAO sources which have fainter flux densities ($S_{\rm 1mm} \lesssim 1$~mJy) may have smaller source sizes. 
This is proved to be valid for the ASAGAO sources based on $uv$-visibility stacking analysis \citep{fuji18}.

\begin{figure}
\begin{center}
\includegraphics[width=\linewidth]{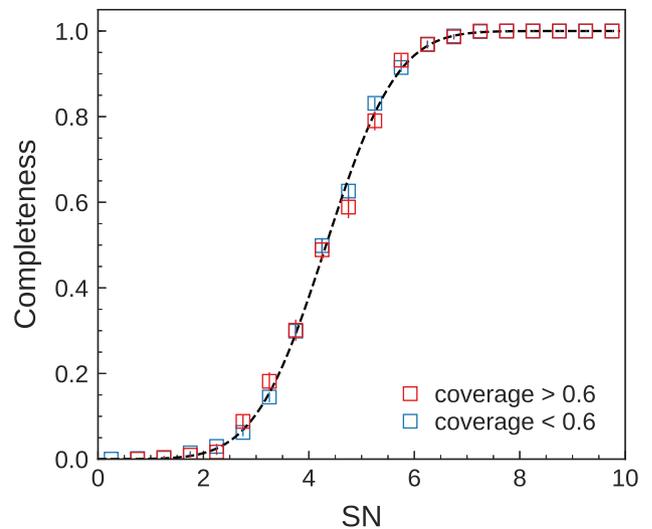}
\end{center}
\caption{
Completeness calculated for the regions with coverage $> 0.6$ (red) and $< 0.6$ (blue) as a function of input peak SN. 
The squares and error bars represent mean and 1$\sigma$ from the binomial distribution within a bin obtained by 30000 trials in each coverage region. 
The dashed curve show the best-fit function of $f({\rm SN}) = [1 +  {\rm erf}(({\rm SN} - a)/b)]/2$ for the entire region, where $(a, b) = (4.33, 1.50)$. 
}
\label{fig:completeness}
\end{figure}

In the completeness calculation, we take a convolving beam size to be uniformly distributed from 0\farcs01--0\farcs5. 
We input 30000 artificial sources into the signal map one at a time, each with an integrated flux density randomly selected from 0.05--2~mJy by considering the flux range of detected sources. 
The input sources are then extracted in the same manner as in Sec.~\ref{subsec:detection}. 
When the input source is detected with a peak SN $\ge 4.5$, the source is considered to be recovered. 
The completeness calculation is conducted separately for the central deeper region (coverage $> 0.6$) and the rest (coverage $< 0.6$) to see the effect of the survey depth. 
The result is shown in figure~\ref{fig:completeness}. 
The completeness calculated in regions with different coverage are consistent within errors and we do not find a significant difference. 
The completeness is 60\% at SN $= 4.5$, and 100\% at SN $\gtrsim 7$.

\begin{figure}
\begin{center}
\includegraphics[width=\linewidth]{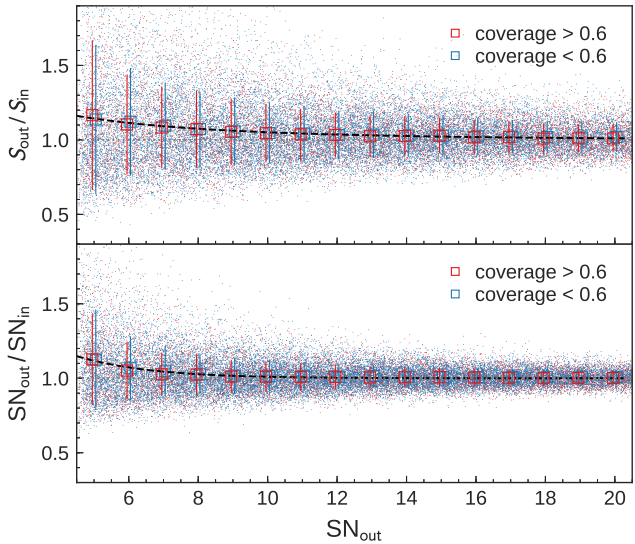}
\end{center}
\caption{
The ratio between input flux ($S_{\rm in}$) and output flux ($S_{\rm out}$) (top) and the ratio between input peak SN (SN$_{\rm in}$) and output peak SN (SN$_{\rm out}$) (bottom) as a function of output peak SN calculated for for the regions with coverage $> 0.6$ (red) and $< 0.6$ (blue).
The 30000 trials in each coverage region are presented as dots. 
The squares and error bars represent mean and 1$\sigma$ with an bin. 
The dashed curves show the best-fit function of $f({\rm SN}) = 1 + \exp(a {\rm SN}^b)$ for the entire region, where $(a,b) = (-0.725, 0.612)$ and $(-0.360, 1.11)$ for the top and bottom panels, respectively. 
}
\label{fig:boosting}
\end{figure}

When dealing with low SN sources, we need to consider the effect that flux densities are boosted by noise \citep{murd73, hogg98}. 
In the course of the completeness simulation, we calculate the ratio between input and output integrated flux density to estimate the intrinsic flux density of the detected sources (figure~\ref{fig:boosting}, top panel). 
The effect of flux boosting for the sources with SN $\ge 4.5$ is on average less than 15\%, and the deboosted flux densities range from 135~$\mu$Jy to 1.97~mJy. 
As in the completeness calculation, we do not see any significant difference in the flux boosting for the different coverage regions. 
The fraction of output peak SN and input peak SN is also calculated and shown in figure~\ref{fig:boosting} (bottom panel).

\subsection{1.2mm Number Counts} \label{subsec:counts}

\begin{figure}
\begin{center}
\includegraphics[width=\linewidth]{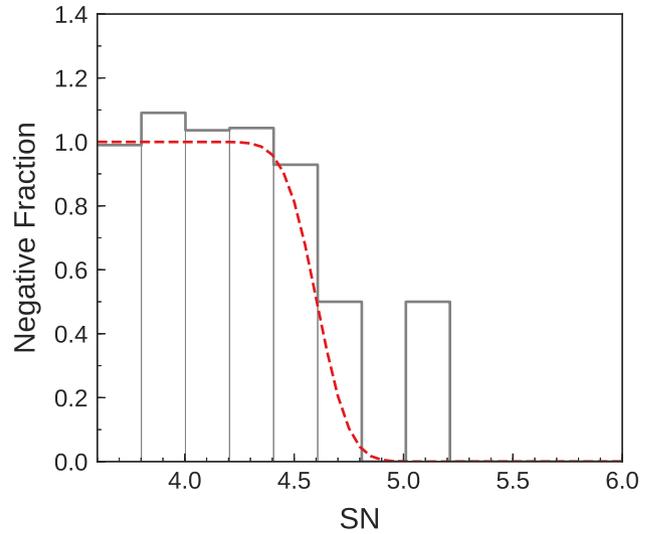}
\end{center}
\caption{
The differential fraction of negative peaks to positive peaks as a function of peak SN. 
The dashed curve represents the best-fit function of $f({\rm SN}) = [1 +  {\rm erf}(({\rm SN} - a)/b)]/2$, where $(a, b) = (4.60, 0.165)$. 
}
\label{fig:fdr_frac}
\end{figure}

\begin{figure*}
\begin{center}
\includegraphics[width=\linewidth]{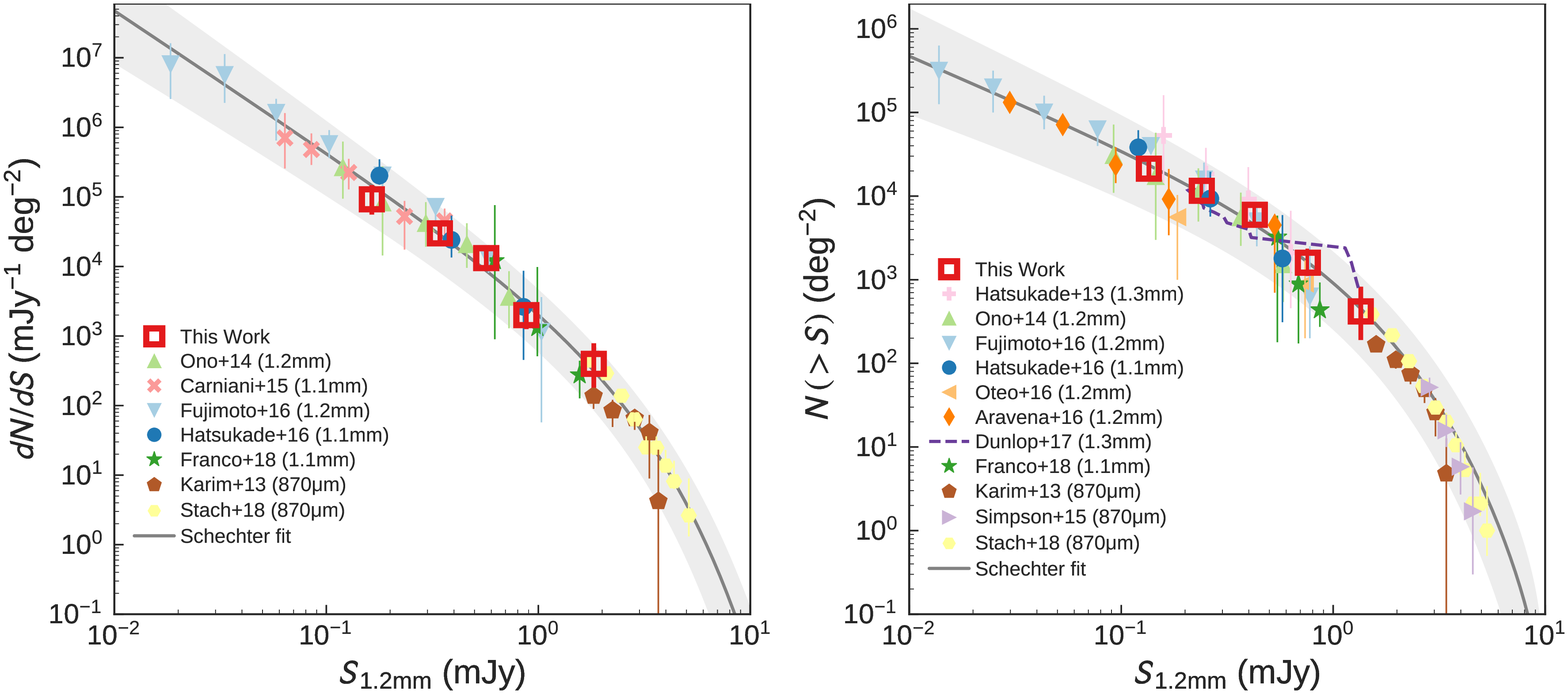}
\end{center}
\caption{
Differential (left) and cumulative (right) number counts at 1.2~mm obtained for ASAGAO sources (red squares). 
For comparison, we plot the results for the 
ALMA blank field surveys of SXDF-ALMA at 1.1~mm \citep{hats16}, ASPECS at 1.2~mm \citep{arav16}, HUDF at 1.3~mm \citep{dunl17}, and GOODS-ALMA at 1.1~mm \citep{fran18}. 
Number counts derived from serendipitously-detected ALMA sources by \citet{hats13}, \citet{ono14}, \citet{carn15}, \citet{fuji16}, and \citet{oteo16} are also presented. 
For the bright end, ALMA 870~$\mu$m follow-up observations of single-dish sources by \citet{kari13}, \citet{simp15b}, and \citet{stac18} are presented. 
The solid curve and shaded area represent the best-fitting functions in the form of Schechter function and 1$\sigma$ error fitted to the differential number counts. 
The flux densities of the counts are scaled to the wavelength of ASAGAO by assuming a modified black body with a dust emissivity index of $\beta = 1.5$, dust temperature of 35~K, and $z = 2$. 
}
\label{fig:counts}
\end{figure*}

By using the 4.5$\sigma$ sources, we create differential and cumulative number counts. 
To create number counts, we correct for the contamination of spurious sources, the effective area, and the completeness as follows: 
\begin{eqnarray}
\frac{dN}{dS} = \frac{1}{\Delta S} \sum_{i} \frac{1 - f_{\rm neg}{({\rm SN}_i})}{A(S_i) C({\rm SN}_i)}, 
\end{eqnarray}
where $S_i$ is the observed source flux density, $f_{\rm neg}$ is the negative fraction accounting for spurious detections, $A$ is the effective area, $C$ is the completeness, and $\Delta S$ is the width of the flux bin. 
Figure~\ref{fig:fdr_frac} shows the differential fraction of the number of negative peaks to positive peaks ($f_{\rm neg}$) as a function of SN. 
The contamination of spurious sources to each source is estimated by using the best-fit function of the negative fraction and is subtracted from unity. 
Then the counts are divided by the completeness by using the best-fit function as a function of SN (figure~\ref{fig:completeness}). 
Here we use SNs corrected for the boosting effect presented in figure~\ref{fig:boosting} (bottom panel). 
The effective area estimated for each flux density is used as the survey area for a source. 
The effect of flux boosting on the source flux density is corrected by using the best-fit function shown in figure~\ref{fig:boosting} (top panel). 
The uncertainties from Poisson fluctuations is estimated from Poisson confidence limits of 84.13\% \citep{gehr86}, which correspond to 1$\sigma$ for Gaussian statistics that can be applied to small number statistics. 
The derived number counts are shown in figure~\ref{fig:counts} and table~\ref{tab:counts}.

The differential number counts obtained in this study and previous studies are fitted to a Schechter function of the form, 
\begin{eqnarray}
\frac{dN}{dS} = \frac{N'}{S'} \left( \frac{S}{S'} \right)^{\alpha} \exp{\left( \frac{-S}{S'} \right)}. 
\end{eqnarray}
In this fit, we use the ALMA number counts plotted in figure~\ref{fig:counts}, which are based on blank-field surveys and serendipitously-detected sources at 1.1--1.3~mm to constrain the faint flux range ($<$1 mJy), and the results of 870-$\mu$m follow-up observations of single dish sources \citep{kari13, stac18} for the bright end by scaling the flux densities from 870-$\mu$m to 1.2~mm. 
Here we assume a modified black body with a dust emissivity index of $\beta = 1.5$, dust temperature of 35~K, and $z = 2$. 
The best-fit parameters are summarized in table~\ref{tab:fit_counts}.

\begin{table}
\tbl{Differential and cumulative number counts. \label{tab:counts}}
{
\footnotesize
\begin{tabular}{cccccc}
\hline
$S$  & $N$ & $dN/dS$ & $S$ & $N$ & $N$($>$$S$) \\
(1)  & (2) & (3)     & (4) & (5) & (6)     \\
(mJy)&     & ($10^2$ mJy$^{-1}$ deg$^{-2}$) & (mJy) & & ($10^2$ deg$^{-2}$) \\
\hline
0.180 &  6 & $924^{+552}_{-366}$    & 0.135 & 45 & $213^{+64}_{-43}$     \\
0.341 & 12 & $299^{+114}_{-85}$     & 0.240 & 39 & $116^{+26}_{-20}$     \\
0.568 & 17 & $132^{+40}_{-32}$      & 0.427 & 27 & $ 60^{+15}_{-12}$     \\
0.878 &  7 & $ 20.0^{+10.7}_{-7.4}$ & 0.759 & 10 & $ 16^{+7.5}_{-4.9}$   \\
1.828 &  3 & $  4.0^{+3.9}_{-2.2}$  & 1.350 &  3 & $  4.2^{+4.1}_{-2.3}$ \\
\hline
\end{tabular}}
\begin{tabnote}
(1) Weighted-mean flux density for bin center. 
(2) Number of sources for differential number counts. 
(3) Differential number counts. 
(4) Flux density for bin minimum. 
(5) Number of sources for cumulative number counts. 
(6) Cumulative number counts. 
\end{tabnote}
\end{table}

\subsection{Comparison with Previous ALMA Studies}\label{subsec:counts_comparison}

We compare the ASAGAO number counts with the previous results in the ALMA blank-field surveys. 
The number counts of SXDF-ALMA are obtained by using 23 (4$\sigma$) sources detected in a 2~arcmin$^2$ area at 1.1~mm \citep{hats16}. 
The ASPECS number counts are derived from 16 (3$\sigma$) sources detected in a deeper 1~arcmin$^2$ survey at 1.2~mm, covering a fainter flux range \citep{arav16}. 
The HUDF number counts are obtained in a 4.5~arcmin$^2$ survey at 1.3~mm \citep{dunl17} by using 16 sources (3.5$\sigma$, $S_{\rm 1.3mm} > 120$~$\mu$Jy) with secure galaxy counterparts. 
The GOODS-ALMA number counts are obtained from 20 sources (4.8$\sigma$) detected in a 69~arcmin$^2$ survey at 1.1~mm \citep{fran18}. 
The ASAGAO number counts are constructed from the largest sample among the blank-field surveys, leading to the small uncertainty from Poisson statistics. 
The flux range connects the fainter range probed by ALMA deep observations and the brighter range constrained by ALMA follow-up observations of single-dish detected sources. 
We find that our number counts are consistent with those of the previous ALMA blank-field surveys. 
The number counts obtained by using the ensemble of serendipitously-detected sources are also compared \citep{hats13, ono14, carn15, fuji16, oteo16}. 
While the faintest bin of \citet{oteo16} is lower than the ASAGAO number counts, these number counts are overall consistent within errors. 
Note that the lower SN thresholds ($\lesssim$4.5--5$\sigma$) adopted in previous studies might include a larger fraction of spurious sources and overestimate the number counts, although the number counts are corrected for the contamination of spurious sources (e.g., \cite{oteo16, hats16, umeh17, umeh18}).

\subsection{Contribution to Extragalactic Background Light} \label{subsec:ebl}

By using the derived differential number counts, we calculate the fraction of the EBL resolved into discrete sources in this survey. 
The integration of the ASAGAO differential number counts yields $7.7^{+1.7}_{-1.2}$ Jy~deg$^{-2}$ ($S_{\rm 1.2mm} > 135$~$\mu$Jy). 
The EBL at 1.2~mm (243~GHz) is estimated from the measurements by the {\sl Planck} satellite \citep{plan14} following \citet{arav16} and \citet{muno17}. 
By interpolating the measurements at 217 and 353 GHz, the EBL at 1.2~mm is calculated to be $15.1 \pm 0.59$ Jy~deg$^{-2}$. 
We find that $52^{+11}_{-8}$\% of the EBL at 1.2~mm is resolved into discrete sources in the ASAGAO map.
The integration of the best-fitting function in the form of Schechter function reaches 100\% at $S_{\rm 1.2mm} \sim 20$~$\mu$Jy, although we note that there is a large uncertainty to extend the function to the faint flux regime. 
The flux density of $\sim$20~$\mu$Jy is comparable to the stacked ALMA 1.3~mm signal ($S_{\rm 1.3mm} = 20.1 \pm 4.6$~$\mu$Jy, corresponding to SFR of $6.0 \pm 1.4$~$M_{\odot}$~yr$^{-1}$) derived by \citet{dunl17} on the positions of 89 galaxies in the redshift range of $1 < z < 3$ and the stellar mass range of $9.3 < \log(M_*/M_{\odot}) < 10.3$. 
This flux density is also comparable to the stacked flux density of 21 NIR sources with 3.6~$\mu$m magnitudes of $m_{\rm 3.6 \mu m} = 22$--$23$ ($S_{\rm 1.1mm} = 29 \pm 15$~$\mu$Jy, corresponding to SFR of several $M_{\odot}$~yr$^{-1}$) in SXDF-ALMA derived by \citet{wang16}, who found that $\sim$80\% of the EBL is recovered by $m_{\rm 3.6 \mu m} < 23$ sources.

To individually detect these faint submm sources, which significantly contribute to the EBL, it is essential to conduct much deeper observations than in existing deep surveys or use gravitational lensing effects. 
\citet{fuji16} showed that nearly 100\% of the EBL can be explained by including gravitational lensed sources at the faint end ($S_{\rm 1.2mm} \sim 20$~$\mu$Jy). 
On the other hand, \citet{muno17} argue that their 1$\sigma$ upper limits to differential counts derived from three galaxy clusters as part of the ALMA Frontier Fields Survey are lower than those of \citet{fuji16} by $\approx$0.5 dex and the resolved fraction is only 32\% down to $S_{\rm 1.1mm} = 13$~$\mu$Jy. 
Since the faintest end of number counts derived from lensed sources depends on the lensing model, deeper surveys in blank fields are essential to resolve this discrepancy.

\begin{table}
\tbl{Best-fit parameters of parametric fit to differential number counts.$^*$ \label{tab:fit_counts}}{
\footnotesize
\begin{tabular}{ccc}
\hline
$N'$                & $S'$  & $\alpha$ \\
($10^2$ deg$^{-2}$) & (mJy) &          \\
\hline
$31.3 \pm 16.6$ & $1.34 \pm 0.30$ & $-2.03 \pm 0.16$ \\
\hline
\end{tabular}}
\begin{tabnote}
$^*$The errors are 1$\sigma$. 
\end{tabnote}
\end{table}

\section{Luminosity Function}\label{sec:lf}

While IR luminosity functions of submm sources have been extensively studied by {\sl Herschel} at wavelengths $\le$ 500~$\mu$m (e.g., grup13, magn13), the results are affected by source blending and sensitivity limit due to the large beam size. 
Studies at 850~$\mu$m--1~mm wavelengths has been very limited \citep{kopr17}. 
In this section, we present the methods of constructing IR LFs from the ASAGAO sources (Sec.~\ref{subsec:lf}), and compare the results with previous studies (Sec.~\ref{subsec:lf_comparison}). 
We estimate the contribution of the ASAGAO sources to the cosmic SFR density (SFRD) at $z \sim 2$ by using the derived LFs (Sec.~\ref{subsec:sfrd}).

\subsection{IR Luminosity Function of ASAGAO Sources}\label{subsec:lf}

\begin{figure}
\begin{center}
\includegraphics[width=\linewidth]{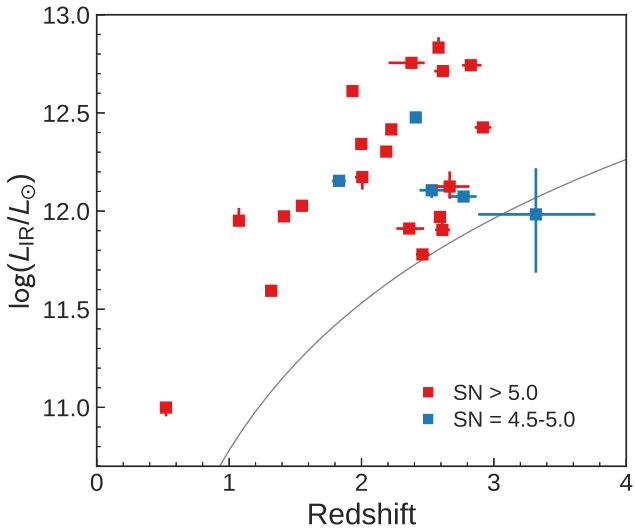}
\end{center}
\caption{
IR luminosity of the ASAGAO sources with $K_S$-band counterpart as a function of redshift. 
The solid curve represents the luminosity limit in this study estimated from the average SED of the 4.5$\sigma$ sources and a detection limit of $\sigma_{\rm 1.2mm} = 26$~$\mu$Jy~beam$^{-1}$. 
}
\label{fig:z-lir}
\end{figure}

To estimate LFs, the redshifts of the ASAGAO sources are required. 
We utilize spectroscopic or photometric redshifts of optical/NIR counterparts. 
We identify $K_S$-band selected sources from the catalog of the {\tt FourStar} galaxy evolution survey (ZFOURGE; \cite{stra16}). 
The ZFOURGE covers a total of 400 arcmin$^2$ including the ASAGAO region with a limiting 5$\sigma$ depth in $K_S$ of 26.0 and 26.3 AB mag for 80\% and 50\% completeness with masking, respectively. 	
The counterpart identification and SED fitting are described in detail in \citet{yama18}, and here we just give a brief explanation. 
The ASAGAO sources are cross-matched with the ZFOURGE catalog. 
For point-like $K_S$-band sources, we adopt a search radius of 0\farcs5, which is small enough to identify a counterpart. 
For extended $K_S$-band sources, we adopt a larger radius, up to half-light radius. 
By using ancillary multi-wavelength data (0.4--500 $\mu$m) and our ALMA photometry, SED fitting with the {\sc magphys} model \citep{dacu08, dacu15} is performed. 
The SED templates of \citet{bruz03} and the dust extinction model of \citet{char00} are adopted.
The number of ASAGAO sources with ZFOURGE counterparts are 20 (80\%) and 25 (56\%) for 5$\sigma$ and 4.5$\sigma$ sources, respectively. 
We use the 5$\sigma$ sources for constructing IR LFs by considering the completeness of the counterpart identification. 
Note that the 5$\sigma$ sources without counterparts are likely to be at higher redshifts ($z \gtrsim$4--5) based on their optical--ratio SEDs \citep{yama18}, and therefore they do not affect the following discussion for the LFs at $z =$ 1--3 significantly. 
The spectroscopic or photometric redshifts are available in the ZFOURGE catalog. 
IR luminosities (measured in the rest-frame 8--1000 $\mu$m) are derived in the SED fitting. 
The IR luminosities as a function of redshift are shown in figure~\ref{fig:z-lir}.

\begin{figure*}
\begin{center}
\includegraphics[width=\linewidth]{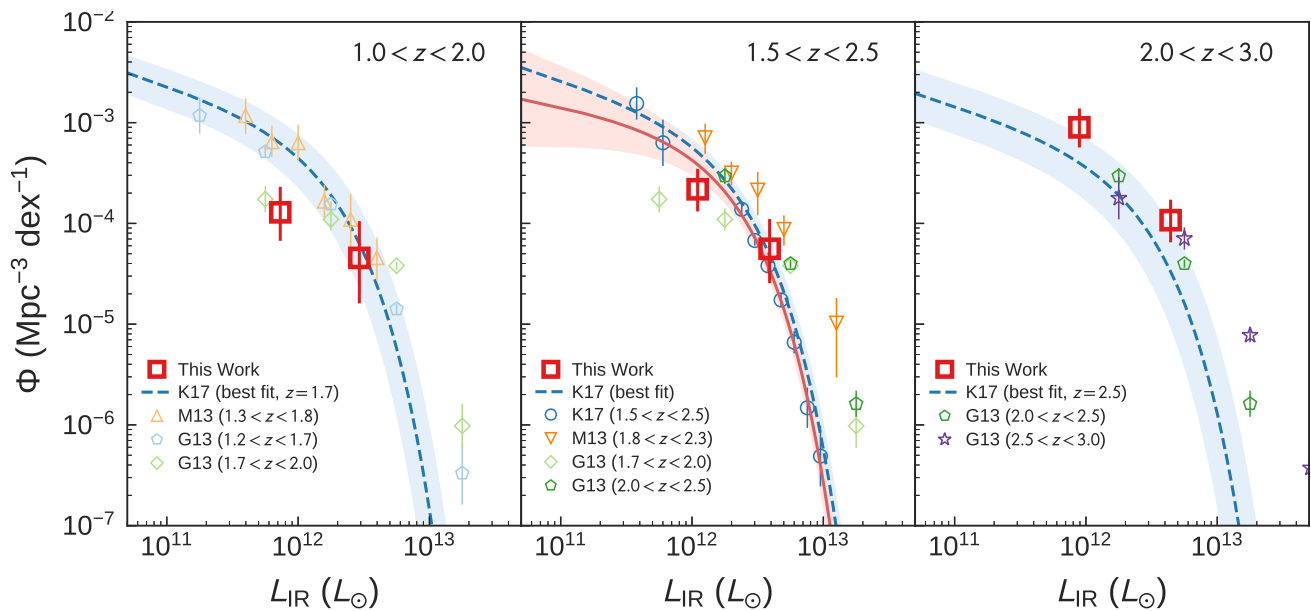}
\end{center}
\caption{
IR luminosity functions constructed from the ASAGAO sources at $1.0 < z < 2.0$ (left), $1.5 < z < 2.5$ (middle), and $2.0 < z < 3.0$ (right).
We plot luminosity functions obtained in \citet{kopr17} (K17) by using 1.3~mm sources from the ALMA HUDF survey and ~$\mu$m sources from the SCUBA-2 Cosmology Legacy Survey. 
The dashed curve and shaded area represent the best-fitting functions and 1$\sigma$ error of \citet{kopr17}. 
At $1.5 < z < 2.5$, we plot their data points derived from the $V_{\rm max}$ method and the best-fitting function. 
At $1.0 < z < 2.0$ and $2.0 < z < 3.0$, we plot their functional form of the redshift evolution of the LF derived from the maximum-likelihood method, adopting the mean redshifts of the ASAGAO sources in each redshift bin ($z = 1.7$ and $z = 2.5$, respectively). 
The results of {\sl Herschel} observations by \citet{magn13} (M13) and \citet{grup13} (G13) are also compared. 
The solid curve and shaded area represent the the best-fitting Schechter function and 1$\sigma$ error fitted to the results of ASAGAO and \citet{kopr17} at $1.5 < z < 2.5$. 
}
\label{fig:lf}
\end{figure*}

To construct the LFs, we adopt the $V_{\rm max}$ method \citep{schm68}. 
This method uses the maximum observable volume of each source. 
The LF gives the number of ALMA sources in a comoving volume per logarithm of luminosity and is obtained as
\begin{eqnarray}
\Phi(L,z) = \frac{1}{\Delta L} \sum_{i} \frac{1}{C({\rm SN}_i) V_{{\rm max},i}}, 
\end{eqnarray}
where $V_{{\rm max},i}$ is the maximum observable volume of the $i$th source, $C$ is the completeness, and $\Delta L$ is the width of the luminosity bin. 
We adopt a luminosity bin width of $\Delta\log{(L)} = 0.6$. 
Because the noise level in the map is not uniform, we need to take into account the effective solid angle where a source can be detected for calculating $V_{\rm max}$. 
Following the description of \citet{nova17}, where they construct radio LFs taking into account a nonuniform noise in their radio maps, we calculate $V_{\rm max}$ as the integration of comoving volume spherical shells as
\begin{eqnarray}
V_{{\rm max},i} = \int_{z_{\rm min}}^{z_{\rm max}} \frac{\Omega(S_i(z))}{4\pi} \frac{dV}{dz} dz, 
\end{eqnarray}
where $z_{\rm min}$ and $z_{\rm max}$ are maximum and minimum redshifts of a redshift bin, $S_i(z)$ is the flux density of source $i$ observed when it is located at $z$, and $\Omega$ is the solid angle where source $i$ with a flux density of $S_i(z)$ can be detected with SN $>$ 5. 
$S_i(z)$ is estimated from the SED model of each source, and $\Omega(S_i(z))$ is derived from the effective area for $S_i(z)$. 
Because the number of sources in each bin is small, the error of the LFs is estimated from Poisson confidence limits of 84.13\% (corresponding to Gaussian 1$\sigma$ errors) in \citet{gehr86}. 
We derive IR LFs in the redshift ranges of $1.0 < z < 2.0$, $1.5 < z < 2.5$, and $2.0 < z < 3.0$ by using 6 (mean redshift of $z_{\rm mean} = 1.55$), 9 ($z_{\rm mean} = 2.12$), and 13 ($z_{\rm mean} = 2.49$) sources, respectively. 
To increase the number of sources in each redshift bin, we adopt the bin width of 1.0, resulting in the overlap of the bins. 
The derived IR LFs are presented in table~\ref{tab:lf} and figure~\ref{fig:lf}. 
Our study constrains the faintest luminosity end of the LF at $2.0 < z < 3.0$ among other studies.

\begin{table}
\tbl{IR Luminosity Functions. \label{tab:lf}}{
\footnotesize
\begin{tabular}{ccc}
\hline
$\log{(L_{\rm IR}/L_{\odot})}^{\dagger}$ & $N$ & $\log{(\Phi/{\rm Mpc^{-3} dex^{-1}})}$ \\
\hline\\
\multicolumn{3}{c}{$1.0 < z < 2.0$} \\
\hline
11.86 & 4 & $-3.89^{+0.25}_{-0.28}$ \\
12.46 & 2 & $-4.34^{+0.37}_{-0.45}$ \\
\hline\\
\multicolumn{3}{c}{$1.5 < z < 2.5$} \\
\hline
11.91 & 6 & $-3.66^{+0.20}_{-0.22}$ \\
12.44 & 3 & $-4.25^{+0.30}_{-0.34}$ \\
\hline\\
\multicolumn{3}{c}{$2.0 < z < 3.0$} \\
\hline
11.94 & 7 & $-3.05^{+0.19}_{-0.20}$ \\
12.57 & 6 & $-3.97^{+0.20}_{-0.22}$ \\
\hline
\end{tabular}}
\begin{tabnote}
$^{\dagger}$ Weighted-mean luminosity in each bin. 
\end{tabnote}
\end{table}

\subsection{Comparison with Previous Studies}\label{subsec:lf_comparison}

We compare the ASAGAO LFs with those derived from sources detected with ALMA, SCUBA2, and {\sl Herschel}. 
\citet{kopr17} derived rest-frame 250~$\mu$m LFs and IR LFs up to $z \sim 5$ by using 16 1.3-mm sources detected in the ALMA HUDF survey \citep{dunl17} for constraining the faint end and 577 850-$\mu$m sources detected in the COSMOS and UDS fields as part of the SCUBA-2 Cosmology Legacy Survey (S2CLS; \cite{geac17, chen16, mich17}) for constraining the bright end. 
The wide coverage of the luminosity range and the large sample for the bright end allowed them to examine the evolution of LFs derived for submm sources. 
They derived LFs for four redshift bins $z =$ 0.5--1.5, 1.5--2.5, 2.5--3.5, and 3.5--4.5, by using the $V_{\rm max}$ method. 
They determined the faint-end slope of $\alpha = -0.4$ in the Schechter form of 
\begin{eqnarray}
\Phi(L) = \Phi_* \left( \frac{L}{L_*} \right)^{\alpha} \exp{\left( \frac{-L}{L_*} \right)}, \label{eq:lf}
\end{eqnarray}
by fitting to the data in the redshift bin of $1.5 < z < 2.5$, where ALMA sources are available for constraining the faint end. 
The remaining Schechter-function parameters were determined by fixing the faint-end slope $\alpha$ to $-0.4$. 
To estimate the continuous form of the redshift evolution of the LF, they used the maximum-likelihood method. 
In figure~\ref{fig:lf}, we plot their data points and the best-fitting function determined in the redshift bin of $1.5 < z < 2.5$, and the LFs determined from the maximum-likelihood method for the redshift bins of $1.0 < z < 2.0$ and $2.0 < z < 3.0$. 
They find that the LFs are well characterized by the number density/luminosity evolution of LFs with positive luminosity evolution coupled with negative density evolution with increasing redshift. 
We find that the ASAGAO LFs are consistent with those of \citet{kopr17} within the errors, supporting the evolution of LFs derived in \citet{kopr17}, although the large uncertainties of our LFs due to the small sample size and the limited coverage of IR luminosity do not allow us to further discuss the density/luminosity evolution of submm sources. 
The ASAGAO LFs at $2.0 < z < 3.0$ is above their results, while those results are consistent. 
This may suggest a stronger luminosity evolution or weaker density evolution. 
The fainter bin of the ASAGAO LFs at $1.0 < z < 2.0$ is about a factor of a few lower than that of \citet{kopr17}. 
This may be due to the fact that they fixed the faint-end slope when deriving the LF evolution.

The results of {\sl Herschel} observations are also compared in figure~\ref{fig:lf}. 
\citet{grup13} derived IR LFs up to $z \sim 4$ by using the data from the {\sl Herschel}-PEP survey in combination with the {\sl Herschel}-HerMES data. 
\citet{magn13} presented IR LFs up to $z \sim 2$ obtained in the GOODS fields from the PEP and the GOODS-{\sl Herschel} programs. 
\citet{kopr17} found that a discrepancy between the results based on submm sources and {\sl Herschel} sources at the bright end, and concluded that {\sl Herschel} results are contaminated and biased high by a mix of source blending, mis-identification of counterpart (and hence redshift) due to the large beam size of {\sl Herschel}/SPIRE. 
Although the {\sl Herschel} results scattered and the redshift ranges are not exactly the same as in ours, we find that they are overall consistent with the ASAGAO LFs.

We fit the IR LFs at $1.5 < z < 2.5$ obtained from the ASAGAO sources and the results of \citet{kopr17} with a Schechter function of the form of equation~\ref{eq:lf}. 
The best-fitting parameters are presented in table~\ref{tab:fit_lf}. 
The derived spectral slope of $\alpha = -0.22 \pm 0.28$ is flatter than $\alpha = -0.4$ derived by \citet{kopr17}, but consistent within the errors. 
In order to constrain the redshift evolution of LFs, it is essential to conduct wider-area surveys for obtaining a larger sample in a wide range of IR luminosity.

\subsection{Contribution to the Cosmic SFR Density}\label{subsec:sfrd}

By integrating the best-fit IR LF and converting it to SFRD, we estimate the contribution of ASAGAO sources to the cosmic SFRD at $z \sim 2$. 
SFR is converted from IR luminosity by using the relation of \citet{kenn98} and corrected to a \citet{chab03} IMF. 
The integration of the best-fitting luminosity function down to the lowest luminosity of the sources ($\log{(L_{\rm IR}/L_{\odot})} = 11.78$) gives a SFRD of $7.2^{+3.0}_{-1.9} \times 10^{-2}$~$M_{\odot}$~yr$^{-1}$~Mpc$^{-3}$. 
This is consistent with the results of \citet{yama18}, where they derived the SFRD by counting the contribution from individual ASAGAO sources. 
We compare the SFRD with the total SFRD (UV $+$ IR) at $z \sim 2$ estimated in previous studies: 
0.13 $M_{\odot}$~yr$^{-1}$~Mpc$^{-3}$ at $z = 2$ by \citet{mada14}, 
or 0.11--0.12 $M_{\odot}$~yr$^{-1}$~Mpc$^{-3}$ at $z =$ 1.8--2.25 by \citet{burg13}. 
The fraction of SFRD contributed by the ASAGAO sources is $\approx$60--90\% at $z \sim 2$, indicating that the major portion of SFRD at that redshift is composed of obscured star formation from sources with $\log{(L_{\rm IR}/L_{\odot})} \gtrsim 11.8$ (e.g., \cite{dunl17, kopr17}). 
This is reasonable considering that the IR luminosity is somewhat lower than the turnover IR luminosity of the best-fit Schechter function.

\begin{table}
\tbl{Best-fit parameters of parametric fit to LF at $1.5 < z < 2.5$ by using the ASAGAO sources and the results of \citet{kopr17}.$^*$ \label{tab:fit_lf}}{
\footnotesize
\begin{tabular}{ccc}
\hline
$\log{(\Phi_*/{\rm Mpc^{-3} dex^{-1}})}$ & $\log{(L_*/L_{\odot})}$ & $\alpha$ \\
\hline
$-3.07 \pm 0.07$ & $12.12 \pm 0.05$ & $-0.22 \pm 0.28$ \\
\hline
\end{tabular}}
\begin{tabnote}
$^*$The errors are 1$\sigma$. 
\end{tabnote}
\end{table}

\section{Conclusions}\label{sec:conclusions}

We performed the ALMA twenty-six arcmin$^2$ survey of GOODS-S at one-millimeter (ASAGAO). 
The central 26~arcmin$^2$ area of the GOODS-S field was observed at 1.2~mm, providing a map with $1\sigma \sim 61$~$\mu$Jy~beam$^{-1}$ (250 k$\lambda$-taper) and a synthesized beam size of $0\farcs51 \times 0\farcs45$. 
By combining the ALMA archival data available in the GOODS-S field (HUDF by \cite{dunl17} and GOODS-ALMA by \cite{fran18}), we obtained a deeper map for the 26~arcmin$^2$ area, which has a rms noise level of $1\sigma \sim 30$~$\mu$Jy~beam$^{-1}$ for the central region with a 250 k$\lambda$-taper and a synthesized beam size of $0\farcs59 \times 0\farcs53$. 
We find 25 sources at 5$\sigma$ and 45 sources at 4.5$\sigma$ in the combined ASAGAO map, providing the largest source catalog among ALMA blank field surveys. 
The flux densities are consistent with those estimated in the other ALMA GOODS-S surveys by considering the difference in observing wavelength.

The larger sample allow us to construct 1.2~mm number counts with smaller uncertainties from Poisson statistics. 
The flux coverage of the number counts connects the fainter range probed by ALMA deep observations and the brighter range constrained by ALMA follow-up observations of single-dish detected sources. 
We find that our number counts are consistent with previous ALMA studies. 
By integrating the derived differential number counts, we find that $52^{+11}_{-8}$\% of the EBL at 1.2~mm is revolved into the discrete sources. 
The integration of the best-fitting function reaches 100\% at $S_{\rm 1.2mm} \sim 20$~$\mu$Jy, although there is a large uncertainty to extend the function to the fainter flux range. 
Deeper surveys are required to individually detect faint submm sources, which significantly contribute to the EBL.

By using the 5$\sigma$ sources, we construct IR LFs in the redshift ranges of $1.0 < z < 2.0$, $1.5 < z < 2.5$, and $2.0 < z < 3.0$. 
Our study constrains the faintest luminosity end of the LF at $2.0 < z < 3.0$ among other studies. 
We find that the ASAGAO LFs are consistent with those of \citet{kopr17}, supporting the evolution of LFs (positive luminosity evolution and negative density evolution with increasing redshift) derived in \citet{kopr17}. 
The integration of the best-fitting LF down to the lowest luminosity of the sources ($\log{(L_{\rm IR}/L_{\odot})} = 11.78$) gives a SFRD of $7.2^{+3.0}_{-1.9} \times 10^{-2}$~$M_{\odot}$~yr$^{-1}$~Mpc$^{-3}$. 
We find that the IR-based star formation of ASAGAO sources contribute to $\approx$60--90\% of the SFRD at $z \sim 2$ derived from UV--IR observation, indicating that the major portion of $z \sim 2$ SFRD is composed of sources with $\log{(L_{\rm IR}/L_{\odot})} \gtrsim 11.8$.

\begin{ack}

We are grateful to Maciej Koprowski for providing the scaling factor of their LFs. 
BH, KK, YT, HU, and YU are supported by JSPS KAKENHI Grant Number 15K17616, 17H06130, and 17K14252. 
RJI acknowledges support from ERC in the form of Advanced Investigator Programme, COSMICISM, 321302. 
This study is supported by the NAOJ ALMA Scientific Research Grant Number 2017-06B and 2018-09B, and by the ALMA Japan Research Grant of NAOJ Chile Observatory, NAOJ-ALMA-190. 
This paper makes use of the following ALMA data: ADS/JAO.ALMA\#2015.1.00098.S, \#2012.1.00173.S, and \#2015.1.00543.S. 
ALMA is a partnership of ESO (representing its member states), NSF (USA) and NINS (Japan), together with NRC (Canada), NSC and ASIAA (Taiwan), and KASI (Republic of Korea), in cooperation with the Republic of Chile. The Joint ALMA Observatory is operated by ESO, AUI/NRAO and NAOJ.

\end{ack}


\end{document}